\documentclass[trackchanges, twocolumn]{aastex701}
\usepackage{booktabs}
\usepackage{rotating}
\usepackage{enumerate}
\usepackage{siunitx}

\newcommand{\oii}{[O\,\textsc{ii}]}
\newcommand{\oiii}{[O\,\textsc{iii}]}
\newcommand{\sii}{[S\,\textsc{ii}]}
\newcommand{\siii}{[S\,\textsc{iii}]}

\newcommand{\nii}{[N\,\textsc{ii}]}
\newcommand{\ariii}{[Ar\,\textsc{iii}]}
\newcommand{\ariv}{[Ar\,\textsc{iv}]}

\newcommand{\cliii}{Cl\,\textsc{iii}}
\newcommand{\cliv}{Cl\,\textsc{iv}}

\newcommand{\hb}{H$\rm\beta$}

\newcommand{\ciii}{[C\,\textsc{iii}], C\,\textsc{iii}]}
\newcommand{\ciiisemi}{C\,\textsc{iii}]}
\newcommand{\civ}{C\,\textsc{iv}}
\newcommand{\hei}{He\,\textsc{i}}
\newcommand{\heii}{He\,\textsc{ii}}
\newcommand{\oiiiuv}{O\,\textsc{iii}]}

\newcommand{\cii}{[C\,\textsc{ii}]~}

\newcommand{\siiii}{Si\,\textsc{iii}]}

\newcommand{\hii}{H\,\textsc{ii}~}
\shortauthors{Arangur{\'e} et al.}

\begin{document}

\title{Investigating the highly ionized gas in UM 461: a reionization era analog}

\author[0009-0004-7070-1782]{Oskar Arangur{\'e}}
\affiliation{Space Telescope Science Institute, 3700 San Martin Drive, Baltimore, MD, 21218, USA}
\affiliation{Universidad Nacional Aut{\'o}noma de M{\'e}xico. Instituto de Astronom{\'i}a. A.P. 106, 22800. Ensenada, B.C., M{\'e}xico}
\email[show]{oarangure@stsci.edu}

\author[0000-0002-5320-2568]{Nimisha Kumari}
\affiliation{AURA for European Space Agency (ESA), ESA Office, Space Telescope Science Institute, 3700 San Martin Drive, Baltimore, MD, 21218, USA}
\email{kumari@stsci.edu}

\author[0000-0003-2685-4488]{Claus Leitherer}
\affiliation{Space Telescope Science Institute, 3700 San Martin Drive, Baltimore, MD, 21218, USA}
\email{leitherer@stsci.edu}

\author[0000-0002-7595-121X]{Joris Witstok}
\affiliation{Cosmic Dawn Center (DAWN), Copenhagen, Denmark}
\affiliation{Niels Bohr Institute, University of Copenhagen, Jagtvej 128, DK-2200,
Copenhagen, Denmark}
\email{joris.witstok@nbi.ku.dk}

\author[0000-0001-5801-6724]{Christophe Morisset}
\affiliation{Instituto de Ciencias F{\'i}sicas, Universidad Nacional Aut{\'o}noma de M{\'e}xico, Av. Universidad s/n,
62210 Cuernavaca, Morelos, M{\'e}xico}
\email{morisset@astro.unam.mx}

\author[0000-0001-8034-7802]{Renske Smit}
\affiliation{Astrophysics Research Institute, Liverpool John Moores University, 146 Brownlow Hill, Liverpool L3 5RF, UK}
\email{R.Smit@ljmu.ac.uk}

\author[0000-0001-8289-3428]{Aida Wofford}
\affiliation{Universidad Nacional Aut{\'o}noma de M{\'e}xico. Instituto de Astronom{\'i}a. A.P. 106, 22800. Ensenada, B.C., M{\'e}xico}
\email{awofford@astro.unam.mx}

%\collaboration{all}{Friends}


\begin{abstract}

We investigate the ionized gas within a local ($z\sim0$) metal-poor blue compact dwarf galaxy, UM~461, by making use of the far-ultraviolet (FUV), near-ultraviolet (NUV), optical, mid-infrared, and far-infrared (FIR) spectroscopy obtained using \textit{Hubble}, the Very Large Telescope (VLT), \textit{Spitzer}, and \textit{Herschel}, respectively. The NUV observations of the brightest star-forming region in UM~461 (R1) show a large equivalent width of the rest frame \ciii\ $\lambda\lambda1907,9$ doublet ($18.46\pm1.03$ \AA), comparable to those of the galaxies from the epoch of reionization (EoR). Together with a high FIR \oiii~88$\mu$m/\cii157$\mu$m ($>1$) line ratio, also characteristic of EoR galaxies, this establishes UM~461 as their local analogue. Furthermore, we detect several highly ionized nebular lines in R1 such as \heii~$\lambda4686$ (ionization potential (IP) $\sim54$ eV), \civ~$\lambda\lambda1548,50$ (IP $\sim$48 eV) and \ariv~$\lambda4740$ (IP $\sim$41 eV), which denote the presence of a hard ionizing radiation field. This could originate from weak-winded WNE stars or a stellar initial mass function with a high ($>100~M_\odot$) upper limit. This analysis shows UV diagnostics using \ciiisemi, \siiii\ and \heii\ perform better than the commonly used optical emission line ratio diagrams in metal-poor galaxies. The spatially-resolved optical data allowed us to establish there are no radial gradients of O/H, N/O, S/O, or Ar/O within this galaxy. This work demonstrates the potential of indirectly identifying highly ionized galaxies with extreme UV characteristics with FIR data, while also unveiling the detailed physical insights offered by local analogues.
\end{abstract}


\section{Introduction \label{sec:introduction}}
\setcounter{footnote}{0}
During the epoch of reionization (EoR), at $6\lesssim z\lesssim10$ \citep{Robertson2015, Cain2025}, our Universe transitioned from a neutral to a primarily ionized state \citep{Gunn1965}. The low-mass star-forming galaxies are generally thought to be the main contributors of ionizing photons which led to the reionization \citep[e.g.,][]{Becker2013,Robertson2022,Dayal2025}, though other sources like active galactic nuclei \citep[AGN; e.g.,][]{Harikane2023,Matthee2024,Maiolino2024} are also proposed as possible contributors.

Before 2022, the rest-frame ultraviolet (UV) emission lines from these primordial galaxies were detected in only a handful of EoR galaxies, as they were too faint and redshifted for the available facilities \citep{Stark2015a,Stark2015b,Stark2017,Hutchison2019}. The James Webb Space Telescope (JWST) has significantly improved the observation of EoR galaxies, permitting for the first time the direct determination of multiple properties of these galaxies, such as electron temperatures \citep[e.g.,][]{Schaerer2022A&A...665L...4S,Marconcini2024MNRAS.533.2488M,Arellano-Cordova2025}, star-formation rates \citep[e.g.,][]{Harikane2023ApJS..265....5H,Robertson2024ApJ...970...31R,Tang2026ApJ..1001...38T}, dust attenuation \citep[e.g.,][]{Ji2025ApJ...988L..69J,Saxena2026MNRAS.548ag808S}, Lyman continuum escape fractions \citep[e.g.,][]{Navarro-Carrera2025} and gas-phase abundances \citep[e.g.,][]{Bunker2023A&A...677A..88B,Nakajima2026Natur.653..363N}. Rest-frame UV datasets of high-z galaxies can be complemented by observations of rest-frame far-infrared (FIR) lines, for example, \oiii\ 88 $\mu$m, one of the principal coolants of the interstellar medium (ISM). Therefore, the simultaneous use of JWST and the FIR capabilities of the Atacama Large Millimeter/submillimeter Array (ALMA) opens an important window in our understanding of high-redshift galaxies \citep[e.g.,][]{Carniani2025,Scholtz2025MNRAS.539.2463S,Schouws2025}. 

Local ($z\sim 0$) star-forming dwarf galaxies have been proposed as tools to understand high-z galaxies as they can be studied with higher spatial resolution and deeper observations \citep{Kunth2000A&ARv..10....1K}. These galaxies may share properties such as low-metallicity environments \citep[e.g.,][]{Senchyna2022}, high star-formation rates \citep[e.g.,][]{Madden2013}, strong nebular emission from highly ionized lines \citep[e.g.,][]{Rhoads2023ApJ...942L..14R,Mingozzi2024}, compact sizes \citep[e.g.,][]{Kim2021ApJ...914....2K}, and escape of ionizing photons \citep[e.g.,][]{Izotov2021}. \citet{Kumari2024} demonstrated the usability of the FIR \oiii\ 88 $\mu$m/\cii 157 $\mu$m ratio as a tool to identify local dwarf galaxies as potential EoR analogues. Using this ratio, they identified Pox 186 as the galaxy with the largest EW([\ciiisemi\ $\lambda 1907+$[\ciiisemi\ $\lambda 1909$) in the local Universe. While most $z\sim0$ galaxies have EW(\ciiisemi) $<5$ \AA\ \citep{Rigby2015ApJ...814L...6R}, Pox 186 possesses a strong EW(\ciiisemi) $=35.85\pm0.73$ \AA. This is similar to the high EWs ($>15$~\AA) found in EoR galaxies \citep[e.g.,][]{Stark2017,Hutchison2019,Topping2021,Tang2025,Kumari2026arXiv260930003K}.

\begin{figure}
    \centering
    \includegraphics[width=\linewidth]{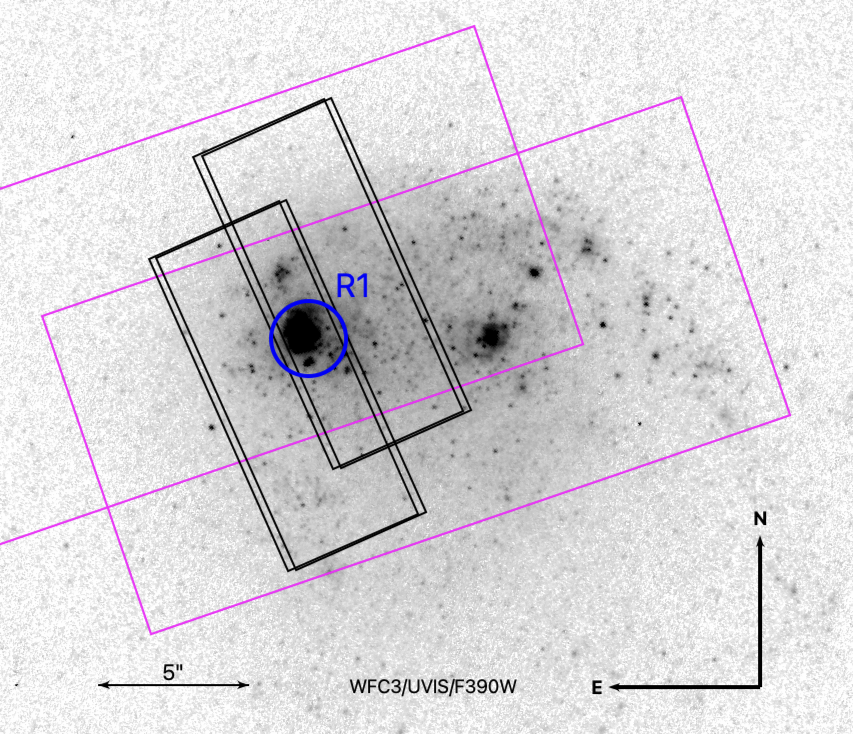}
    \caption{HST/WFC3 image of UM 461 in the F390W filter (PID 15860, PI Elmegreen). The blue circle represents the 2.5 arcsec aperture of HST/COS on the targeted brightest star-forming region, R1. The HST/COS far-ultraviolet and near-ultraviolet spectroscopy of R1 was obtained as part of Program PID: 17729 (PI: Kumari). The \textit{short-high} and \textit{long-high} apertures of the \textit{Spitzer} mid-infrared spectroscopy (PID 85, PI Houck) are presented as black and magenta boxes, respectively.}
    \label{fig:F390W_UM461}
\end{figure}

This paper is the second in a series of papers studying potential EoR analogues identified on the basis of their relatively high FIR \oiii/\cii line ratio. We present a multi-wavelength analysis of UM 461, a blue compact dwarf (BCD) galaxy with a high \oiii\ 88 $\mu$m/\cii 157 $\mu$m ratio \citep[$3.6\pm0.1$,][]{Cormier2015}. We aim to characterize this potential EoR analogue and test different diagnostics that can be applied to high-z galaxies. In Table~\ref{tab:data}, we list the properties of UM 461 found in the literature and in the Mikulski Archive for Space Telescopes (MAST). This galaxy shares similarities with Pox 186 as they are local ($z\sim 0$) blue compact dwarfs with low gas-phase metallicities ($12+\log($O/H$)\sim7.8$) and active star formation. Unlike in Pox 186, there are multiple regions of active star formation within UM 461. Figure \ref{fig:F390W_UM461} shows the Hubble Space Telescope (HST) WFC3/F390W image of UM 461 taken as part of PID 15860 (PI Elmegreen), where we have marked the brightest star-forming region as R1.

The paper is organized as follows. In Section~\ref{sec:observations}, we describe the far-ultraviolet (FUV) near-ultraviolet (NUV), optical, mid-infrared (MIR), and far-infrared observations of UM 461 used in this analysis. In Section~\ref{sec:results}, we describe the methodology to estimate emission line fluxes, dust-correction, emission line ratios and discuss the properties of UM 461 with respect to other low-metallicity galaxies from literature via various diagnostic diagrams involving UV and optical emission line ratios that can be used in the analysis of high-z galaxies. We also compare the line profiles of the observed UV lines. In Section~\ref{sec:discussion}, we discuss the potential sources of He$^+$ ionizing photons, the observed projected abundance patterns in UM 461, and compare multiple methods suggested in the literature to calculate the escape of ionizing photons. Finally, in Section~\ref{sec:conclusions}, we present our conclusions. 

Throughout this paper, we consider the solar total metallicity by mass as $Z_\odot=0.014$ and a solar gas-phase oxygen abundance of $12+\log$(O/H)$_\odot = 8.69$ \citep{Asplund2021}. We assume a standard flat $\Lambda$CDM cosmology from Planck18 \citep[$H_0=67.8$ km s$^{-1}$ Mpc$^{-1}$, $\Omega_m=0.315$]{Planck2020}.

\begin{table}
\centering
\small
\caption{Literature data and HST/COS exposure times for UM 461}
\resizebox{1.02\columnwidth}{!}
{%
\begin{tabular}{@{}lcr@{}}
\toprule
& \multicolumn{1}{c}{UM 461}                    \\ 
\midrule
RA (J2000)            &     & 11$^\mathrm{h}$51$^\mathrm{m}$33$^\mathrm{s}$.354      \\
Dec (J2000)           &     & \ang[angle-symbol-over-decimal]{ -02;22;21.89}      \\
Redshift$^a$          &     & $0.003465\pm0.000006$         \\
M$_\star$ [10$^7$ M$_\odot$]$^b$          &     & 4.8         \\
%12 + log(O/H)$^c$     &     & $7.81\pm0.02$ \\
Metallicity ($Z$)     &     & 0.13 $Z_\odot$ \\
$E(B-V)_\mathrm{Gal}$$^c$ &     & $0.0148\pm0.0010$ \\
\midrule
COS Setting (Epoch)                & Central wavelength [\AA] & Exposure time [s] \\
\midrule
G160M$^d$                 & 1623 & 3944.512 \\
G185M$^d$ (2025 February 24)                 & 1900 & 7518.560 \\
G185M$^d$ (2026 January 9)                 & 1900 & 9817.152 \\
G185M$^d$ (2026 January 11)                 & 1900 & 2364.576 \\
Mirror B & - & $2\times16$ \\
\bottomrule
\end{tabular}%
}
\\\textit{Notes}.
$^a$\citet{Comte1999};
$^b$\citet{Madden2014};
%$^c$\citet{Izotov2006};
$^c$\citet{Schlafly2011};
$^d$\textit{HST}/COS PID 17729.
\label{tab:data}
\end{table}

\section{Observations}
\label{sec:observations}

\subsection{HST/COS UV spectroscopy}

\begin{figure}
    \centering
    \includegraphics[width=\linewidth]{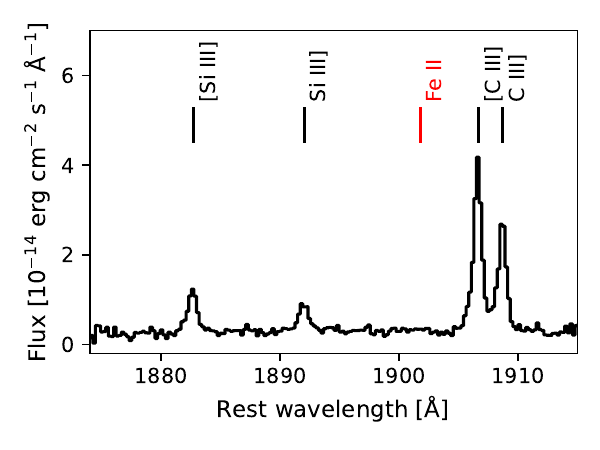}
    \caption{\textit{HST}/COS G185M/1913 spectrum of UM 461 R1. The UV nebular (black) and ISM (red) emission lines from \citet{Leitherer2011} are indicated. The spectrum is rebinned by 6 pixels. The first and last segments of the grating are not shown as they do not have any detected spectral features.
    }
    \label{fig:COS_G185M_um461}
\end{figure}

\begin{figure*}
    \centering
    \includegraphics[width=\linewidth]{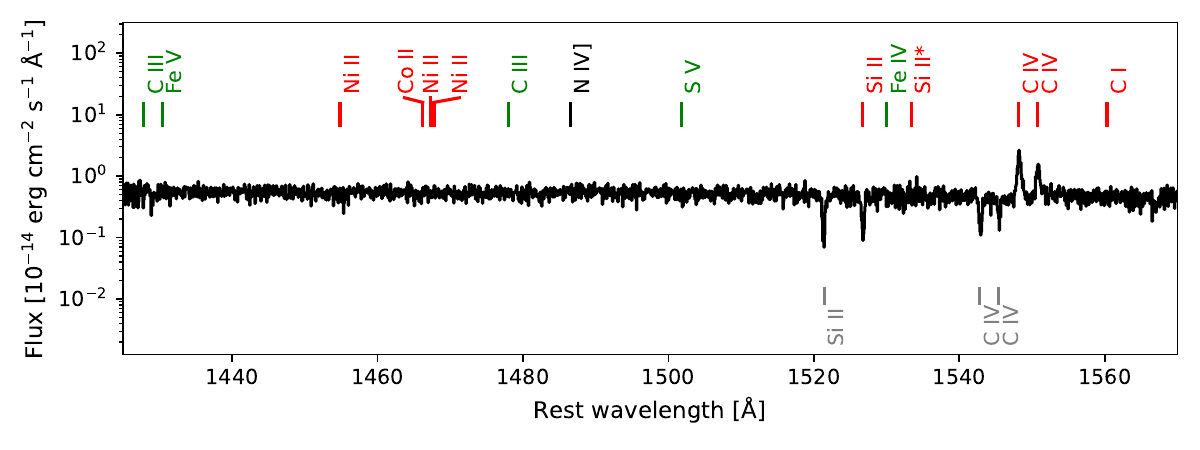}
    \includegraphics[width=\linewidth]{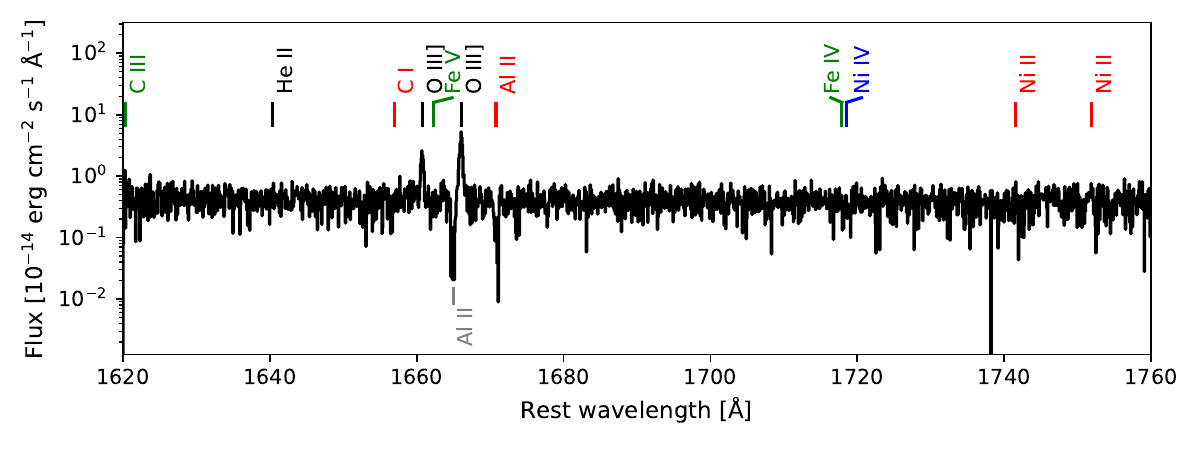}
    \caption{\textit{HST}/COS G160M/1623 spectra of UM 461 R1. UV nebular (black), ISM (red), photospheric (green), and wind (blue) emission lines from \citet{Leitherer2011} are marked. Galactic absorption lines are marked in gray. Not all displayed lines are detected; they are shown for reference only. The spectrum is rebinned by 6 pixels.}
    \label{fig:COS_G160M_um461}
\end{figure*}

The HST/COS FUV and NUV spectroscopy of the brightest star cluster in UM 461 (R1) was obtained as part of the General Observers program in HST Cycle 32 (GO: 17729, PI: Kumari). The blue circle on Figure \ref{fig:F390W_UM461} shows the location of the 2.5 arcsec diameter Primary Science Aperture (PSA) centered at R1. The FUV and NUV data were acquired with the medium resolution gratings G160M and G185M centered at 1623 and 1900 \AA, respectively. On visit 3 (2025 February 24) a failure occurred, resulting in no mirror B acquisition image being obtained; consequently, it could not be ensured that the target was centered for the spectroscopy. Visit 3 observations were retaken, confirming the target was indeed centered. We consider the G185M data from the original and repeat observations to increase the signal-to-noise (S/N). We combined and reduced all the data using the standard data reduction pipeline CALCOS (version 3.6.1).

We rebin the FUV and NUV spectra by 6 pixels using the \textsc{SpectRes} \citep{Carnall2017arXiv} \textsc{python} package. Figures~\ref{fig:COS_G185M_um461} and \ref{fig:COS_G160M_um461} show the resulting NUV and FUV spectra, respectively, along with the location of the UV nebular (black), interstellar medium (red), photospheric (green), and wind (blue) emission lines taken from \citet{Leitherer2011}.

\subsection{Optical spectroscopy}

The optical integral field spectroscopy data of UM 461 were taken with the Multi-Unit Spectroscopic Explorer \citep[MUSE;][]{Bacon2010} on the Very Large Telescope (VLT) as part of program 0104.D-0199 (PI: Brinchmann). This observation covers the wavelength range of 4600-9352 \AA\ with spectral resolving power of $R\sim1770-3590$. The total integration time is 1800 s. The observation was acquired using adaptive optics \citep{Arsenault2008,Strobele2012} in WFM-AO-E mode, which required a Na filter to avoid contamination from the laser guide stars. Due to this, the range of 576-601 nm is unavailable in the spectra\footnote{See the MUSE user manual: \url{https://www.eso.org/sci/facilities/paranal/instruments/muse/doc/ESO-261650_MUSE_User_Manual.pdf}}. The pixel sampling is $0.2^"\times0.2^"$. The data were reduced using the MUSE data reduction pipeline \citep{Weilbacher2020} version 2.8.1. The sky coordinates of UM 461 in the Integral Field Unit (IFU) data from ESO archive are offset with respect to the reported values in the literature. We correct the astrometric offset of IFU data cube by first generating an image from the MUSE data cube corresponding to the HST/WFC3 F390W bandpass, which allowed us to compare and match the positions of the two brightest star-forming regions of UM 461 to those of the actual F390W image. 

\begin{figure*}
    \centering
    \includegraphics[width=\linewidth]{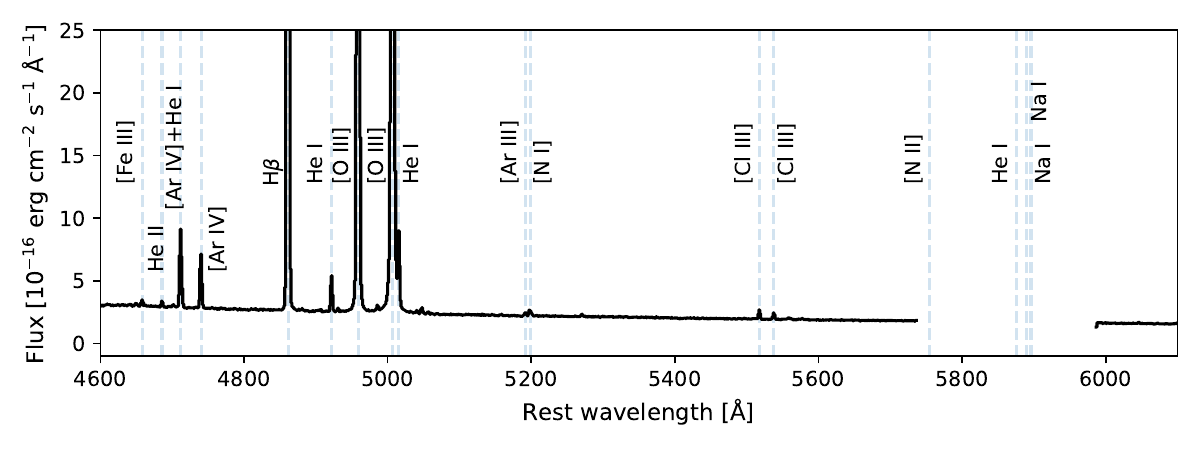}
    \includegraphics[width=\linewidth]{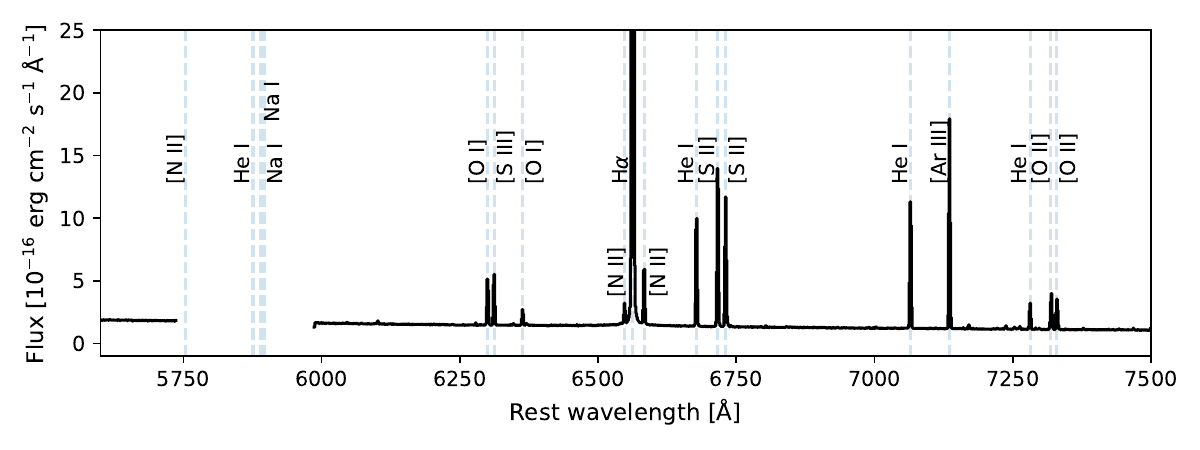}
    \includegraphics[width=\linewidth]{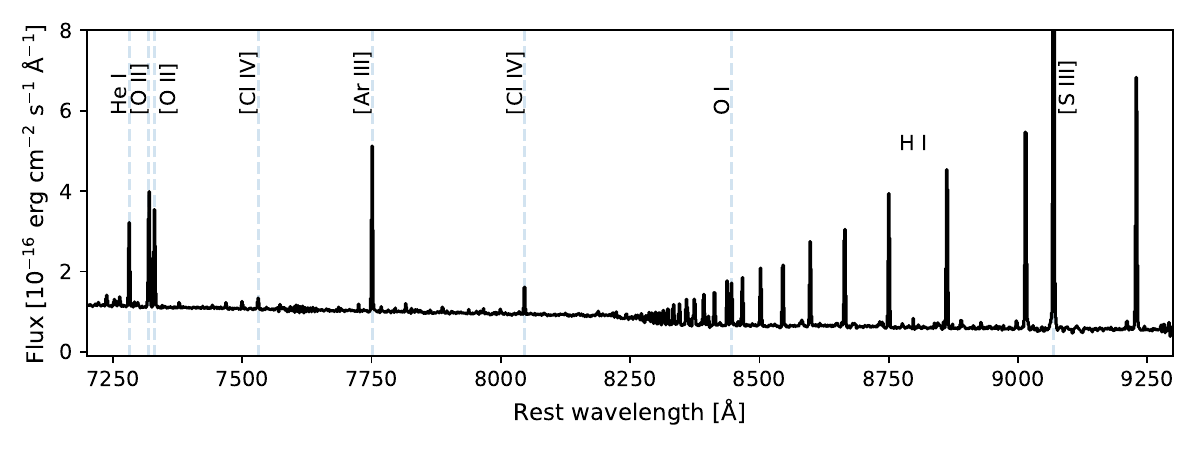}
    \caption{Optical spectra of UM 461 R1 in the wavelength range of 4600-6100 \AA\ (\textit{top panel}), 5600-7500 \AA\ (\textit{middle panel}), and  7200-9300 \AA\ (\textit{bottom panel}). The wavelength gap in the middle panel is due to the Na filter, used to avoid contamination from the laser guide stars in the MUSE WFM-AO-E mode. Observed lines are marked with dashed lines. We also mark the location of the Na~\textsc{i} $\lambda\lambda 5890,6$ doublet, He~\textsc{i} $\lambda 5876$, and the \nii\ $\lambda 5755$ auroral line in the region masked by the Na filter.}
    \label{fig:Muse_um461_r1}
\end{figure*}

\indent In order to study region R1 across the UV and optical wavelength range, we integrate the MUSE IFU spectra within a circular aperture matching the COS aperture (diameter of 2.5 arcsec) centered on the region. The resulting integrated spectrum is shown in Figure~\ref{fig:Muse_um461_r1}.

\begin{figure}
    \includegraphics[width=\linewidth]{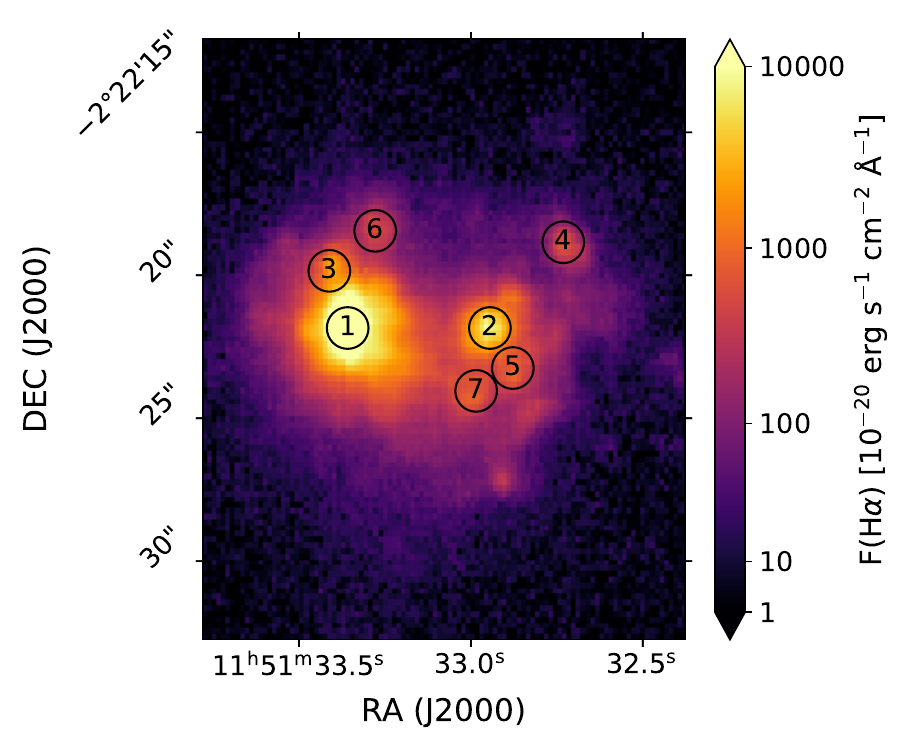}
    \caption{Star-forming regions identified on the VLT/MUSE H$\alpha$ map. North is up, East to the left. We show only the region surrounding UM 461 for better visibility. The sky coverage of the complete VLT/MUSE cube is 1.1 arcmin$^2$.}
    \label{fig:MUSE_Ha}
\end{figure}

In addition, we use MUSE data to identify individual H~\textsc{ii} regions to study variations in chemical abundances and ratios within UM 461. For doing so, we use the publicly available \texttt{ifuanal} package \citep{Lyman2018}, where the identification and extraction of the regions is done with a \texttt{python} implementation of the \texttt{HIIexplorer}\footnote{\url{http://www.caha.es/sanchez/HII_explorer/}} algorithm \citep{Sanchez2012}. This implementation uses the \texttt{find\_peaks} function in \texttt{SciPy} \citep{Virtanen2020} to find maxima at a user-defined wavelength in the cube and then searches for surrounding spaxels brighter than a predefined threshold within a user-defined radius. To find the peaks, we set a minimum H$\alpha$ flux of $3\times 10^{-17}$ erg s$^{-1}$ cm$^{-2}$ \AA$^{-1}$. Spaxels within a radius of 5 pixels ($\sim$1 arcsec) from the peak location and a flux $\ge 10\%$ of its H$\alpha$ flux are selected as part of a region. We identify and extract the spectra of the 7 star-forming clusters shown in Figure~\ref{fig:MUSE_Ha}. The regions are numbered in ascending order of equivalent width of H$\alpha$. R1 is identified as region 1 by \texttt{ifuanal}. For the rest of this paper, we refer to the COS aperture-matched data as R1 and to the \texttt{ifuanal}-defined region simply as region 1. Lowering the minimum peak flux increases the number of identified regions; however, none of these extra regions present detectable emission of the \ariii\ $\lambda$5192 nor \siii\ $\lambda$6312 auroral lines, required for the estimation of ionic temperatures. Using a smaller radius of 4 pixels, representative of the observational seeing of the cube ($\sim$0.74 arcsec), does not change the number of identified regions. 
\citet{Elmegreen2022} identified 6 main clumps based on their H$\alpha$ photometry (see the lower-right panel in their Figure 15). Their clumps spatially agree with our spectroscopically selected clusters. We note that regions 5 and 7 in this study are included within their larger B clump.

\subsection{Ancillary infrared spectroscopy}
\underline{MIR:} UM 461 was observed by the \textit{Spitzer} Space Telescope with the Infrared Spectrograph \citep[IRS;][]{Houck2004}, using both the low ($R\sim64-128$) and high ($R\sim 600$) resolution modules. \citet{Cormier2015} lists UM 461 as a compact source, and extracts the flux of R1 using the CASSIS high-resolution pipeline \citep{Lebouteiller2015} by selecting a spatial window that depends on the extent of the source and wavelength. In this work, we consider MIR fluxes of [S~\textsc{iv}] 10.5 $\mu$m, [Ne~\textsc{ii}] 12.8 $\mu$m, [Ne~\textsc{iii}] 15.6 $\mu$m, [S~\textsc{iii}] 18.7 $\mu$m, and [S~\textsc{iii}] 33.5 $\mu$m obtained from the high resolution IRS data (AOR key 16204032).

\underline{FIR:} UM 461 was observed by the \textit{Herschel} Space observatory \citep{Pilbratt2010} with the Photodetector Array Camera and Spectrometer \citep[PACS][]{Poglitsch2010} with a beam size of 9$''$ and 12$''$ at 60 $\mu$m and 160 $\mu$m, respectively. \citet{Cormier2015} extracted the emission line fluxes by combining the 5 $\times$ 5 spaxels of the PACS array and applying a 5 $\times$ 5 point-source correction since the source is largely compact. In this work, we use the fluxes of the \oiii\ 88 $\mu$m and [C~\textsc{ii}] 157 $\mu$m FIR emission lines. Although these observations cover the whole galaxy, the distribution of flux around the coordinates of R1 shows this region is the dominant contributor of these lines.

\section{Results}
\label{sec:results}

\subsection{UM 461 redshift and distance} \label{sec:distance}
We estimate the redshift of UM 461 from the median of the fitted central wavelengths (see Section~\ref{sec:line_fitting}) of the optical emission lines presented in Table~\ref{tab:um461_fluxes}. We do not use the blended optical lines. The observed redshift of UM 461, $z=0.00344\pm 0.00007$, agrees with the value from the NASA Extragalactic Database (NED, Table~\ref{tab:data}). Considering the Planck18 cosmology \citep{Planck2020}, this redshift corresponds to a luminosity distance of $15.3\pm0.3$ Mpc. The angular scale at this distance is $74\pm1$ pc arcsec$^{-1}$. Cosmicflows-3 \citep{Kourkchi2020} yields a similar distance of 15.7 Mpc.

\subsection{Emission line fluxes \label{sec:line_fitting}}

We perform a linear fit to the local continuum from each emission line under study and subtract it from the spectrum before fitting emission lines. Optical fluxes were obtained by fitting a Gaussian profile to each emission line using the non-linear least-squares minimization (Levenberg–Marquardt algorithm). The velocity dispersions of the lines are not tied due to the variable spectral resolution of MUSE across the broad wavelength range covered and the velocities agree within 15 km s$^{-1}$ for resolved lines, and within 40 km s$^{-1}$ for the blended lines listed in the footnote of Table~\ref{tab:um461_fluxes}. For the UV lines, we directly integrate the flux under the emission line due to a potential underlying broad component in the \ciiisemi\ doublet. This extra component will be further analyzed in future work. Table~\ref{tab:um461_fluxes} lists the observed (F$_\lambda$) and intrinsic (I$_\lambda$) fluxes relative to H$\beta$ of R1. 

We correct the spectrum for foreground Galactic extinction using the attenuation law of \citet{Cardelli1989} through the \texttt{dust\_extinction} python package \citep{Gordon2024}. Galactic $E(B-V)$ in the direction of UM 461 is $0.0148\pm0.0010$ mag \citep{Schlafly2011}. For intrinsic extinction, we assume the Small Magellanic Cloud (SMC) attenuation law of \citet{Gordon2003} with $R_V = 3.41$ mag, given the sub-solar metallicities of UM 461 and the SMC. The intrinsic color excess values ($E(B-V)_{\mathrm{intrinsic}}$) were calculated using the Balmer decrement from the H$\alpha/$H$\beta$ ratio, considering an intrinsic value of 2.78 by assuming Case B recombination for an electron temperature of 16000 K and density of 100 cm$^{-3}$, with the H$^0$ transition probabilities of \citet{Storey1995}. The uncertainties of the continuum fitting, flux measurement and reddening corrections were propagated in quadrature to the errors of the emission line fluxes.

The final $E(B-V)_{\mathrm{intrinsic}}$ and measured optical fluxes of the COS-matched aperture are reported in Table~\ref{tab:um461_fluxes}. The respective values of the individual H~\textsc{ii} regions are included in Appendix~\ref{sec:individual_abundances} (Tables~\ref{tab:individual_hii_region_fluxes} and \ref{tab:individual_hii_region_fluxes_2}).

% The brackets in the names of the elements produce errors in latex.
\begin{deluxetable*}{lcccccc} 
%\digitalasset
\tablewidth{0pt}
\tablecaption{UM 461 optical emission line fluxes (relative to H$\beta$=100) of the COS-matched aperture, R1. \label{tab:um461_fluxes}}
\tablehead{
\colhead{Line} & \colhead{Rest wavelength [\AA]} & \colhead{IP [eV]} & \colhead{F$_\lambda$ (R1)} & \colhead{I$_\lambda$ (R1)}
}
\startdata
$[$Fe~\textsc{iii}$]$ & 4658.1 & 16.20 & 0.159 $\pm$ 0.036 & 0.161 $\pm$ 0.036 \\
He~\textsc{ii} & 4685.68 & 54.42 & 0.131 $\pm$ 0.020 & 0.132 $\pm$ 0.021 \\
$[$Ar~\textsc{iv}$]+$He~\textsc{i} & 4712.0$^a$ & 40.73 & 2.390 $\pm$ 0.068 & 2.412 $\pm$ 0.068 \\
$[$Ar~\textsc{iv}$]$ & 4740.2 & 40.73 & 1.443 $\pm$ 0.025 & 1.453 $\pm$ 0.025 \\
H$\beta$ & 4861.36 & 13.60 & 100.0 $\pm$ 1.5 & 100.0 $\pm$ 1.5 \\
He~\textsc{i} & 4921.93 & 24.59 & 0.909 $\pm$ 0.048 & 0.906 $\pm$ 0.048 \\
$[$O~\textsc{iii}$]$ & 4958.91 & 35.12 & 216.0 $\pm$ 2.5 & 214.7 $\pm$ 2.4 \\
$[$O~\textsc{iii}$]$ & 5006.84 & 35.12 & 595.7 $\pm$ 17.8 & 590.6 $\pm$ 17.6 \\
$[$Ar~\textsc{iii}$]$ & 5191.82 & 27.63 & 0.086 $\pm$ 0.007 & 0.084 $\pm$ 0.007 \\
$[$N~\textsc{i}$]$ & 5199.0$^b$ & 0.00 & 0.203 $\pm$ 0.055 & 0.199 $\pm$ 0.053 \\
$[$Cl~\textsc{iii}$]$ & 5517.71 & 23.81 & 0.241 $\pm$ 0.010 & 0.233 $\pm$ 0.010 \\
$[$Cl~\textsc{iii}$]$ & 5537.87 & 23.81 & 0.168 $\pm$ 0.010 & 0.163 $\pm$ 0.009 \\
$[$O~\textsc{i}$]$ & 6300.3 & 0.00 & 1.298 $\pm$ 0.087 & 1.221 $\pm$ 0.082 \\
$[$S~\textsc{iii}$]$ & 6312.06 & 23.34 & 1.434 $\pm$ 0.089 & 1.348 $\pm$ 0.084 \\
$[$O~\textsc{i}$]$ & 6363.78 & 0.00 & 0.436 $\pm$ 0.027 & 0.409 $\pm$ 0.025 \\
$[$N~\textsc{ii}$]$ & 6548.05 & 14.53 & 0.640 $\pm$ 0.051 & 0.597 $\pm$ 0.047 \\
H$\alpha$ & 6562.85 & 13.60 & 298.3 $\pm$ 6.6 & 278.3 $\pm$ 6.1 \\
$[$N~\textsc{ii}$]$ & 6583.45 & 14.53 & 1.755 $\pm$ 0.061 & 1.637 $\pm$ 0.057 \\
He~\textsc{i} & 6678.15 & 24.59 & 3.254 $\pm$ 0.044 & 3.027 $\pm$ 0.040 \\
$[$S~\textsc{ii}$]$ & 6716.44 & 10.36 & 4.743 $\pm$ 0.056 & 4.407 $\pm$ 0.052 \\
$[$S~\textsc{ii}$]$ & 6730.82 & 10.36 & 3.654 $\pm$ 0.062 & 3.393 $\pm$ 0.058 \\
He~\textsc{i} & 7065.19 & 24.59 & 3.690 $\pm$ 0.050 & 3.396 $\pm$ 0.046 \\
$[$Ar~\textsc{iii}$]$ & 7135.8 & 27.63 & 5.621 $\pm$ 0.080 & 5.163 $\pm$ 0.073 \\
He~\textsc{i} & 7281.35 & 24.59 & 0.712 $\pm$ 0.017 & 0.652 $\pm$ 0.016 \\
$[$O~\textsc{ii}$]$ & 7322.0$^c$ & 13.62 & 1.074 $\pm$ 0.202 & 0.982 $\pm$ 0.185 \\
$[$O~\textsc{ii}$]$ & 7332.0$^d$ & 13.62 & 0.899 $\pm$ 0.451 & 0.822 $\pm$ 0.412 \\
$[$Cl~\textsc{iv}$]$ & 7530.54 & 39.80 & 0.094 $\pm$ 0.011 & 0.086 $\pm$ 0.010 \\
$[$Ar~\textsc{iii}$]$ & 7751.06 & 27.63 & 1.385 $\pm$ 0.021 & 1.254 $\pm$ 0.019 \\
$[$Cl~\textsc{iv}$]$ & 8045.62 & 39.80 & 0.257 $\pm$ 0.010 & 0.231 $\pm$ 0.009 \\
$[$S~\textsc{iii}$]$ & 9069.0 & 23.34 & 8.551 $\pm$ 0.119 & 7.537 $\pm$ 0.105 \\
\midrule
F(H$\beta)^e$ & - & - & 72.49 $\pm$ 0.761 & 88.96 $\pm$ 0.934  \\
$E(B-V)_\mathrm{intrinsic}$ & - & - & 0.068 $\pm$ 0.022 & -  \\
\enddata
\tablecomments{$^a$\ariv\ $\lambda4711$ + He~\textsc{i} $\lambda 4713$. $^b$[N~\textsc{i}] $\lambda$5198 + [N~\textsc{i}] $\lambda$5200. $^c$[O~\textsc{ii}] $\lambda$7319 + [O~\textsc{ii}] $\lambda$7320. $^d$[O~\textsc{ii}] $\lambda$7330 + [O~\textsc{ii}] $\lambda$7331. $^e$F(H$\beta$) in units of $10^{-15}$ erg s$^{-1}$ cm$^{-2}$.}
\end{deluxetable*}

\begin{deluxetable*}{lccccc}
%\digitalasset
\tablewidth{0pt}
\tablecaption{Equivalent widths and emission line fluxes of the UV emission lines. \label{tab:uv_lines}}
\tablehead{
\colhead{Line} & Rest wavelength [\AA] & IP [eV] & \colhead{EW $[$\AA$]$} & \colhead{F$_\lambda$$^a$} & \colhead{I$_\lambda$$^a$}
}
\startdata
C~\textsc{iv} & $1548.19$ & 47.89 & 1.93 $\pm$ 0.19 & 0.42 $\pm$ 0.03 & 0.92 $\pm$ 0.06 \\
C~\textsc{iv} & 1550.77 & 47.89 & 1.27 $\pm$ 0.18 & 0.25 $\pm$ 0.02 & 0.54 $\pm$ 0.05 \\
O~\textsc{iii}$]$ & 1660.81 & 35.12 & 1.82 $\pm$ 0.43 & 0.37 $\pm$ 0.03 & 0.76 $\pm$ 0.06 \\
O~\textsc{iii}$]$ & 1666.15 & 35.12 & 4.55 $\pm$ 0.59 & 0.90 $\pm$ 0.06 & 1.85 $\pm$ 0.12 \\
$[$Si~\textsc{iii}$]$ & 1882.71$^b$ & 16.35 & 4.02 $\pm$ 0.55 & 0.51 $\pm$ 0.04 & 0.96 $\pm$ 0.07 \\
Si~\textsc{iii}$]$ & 1892.03$^b$ & 16.35 & 2.25 $\pm$ 0.31 & 0.34 $\pm$ 0.03 & 0.66 $\pm$ 0.06 \\
\ciiisemi\  & 1908.00$^c$ & 24.38 & 18.46 $\pm$ 1.03 & 3.09 $\pm$ 0.05 & 5.73 $\pm$ 0.09
\enddata
\tablecomments{$^a$In units of $10^{-14}$ erg s$^{-1}$ cm$^{-2}$. $^b$In this work, we use \siiii\ $\lambda\lambda 1888$ to refer to the doublet [Si~\textsc{iii}] $\lambda1883$ and Si~\textsc{iii}] $\lambda1892$. $^c$[\ciiisemi\ $\lambda 1906.68$ + \ciiisemi\ $\lambda 1908.73$. In this work, we use \ciiisemi\ $\lambda\lambda 1908$ to refer to the doublet [\ciiisemi\ $\lambda 1906.68$ and \ciiisemi\ $\lambda 1908.73$.}
\end{deluxetable*}

\subsection{Physical conditions} \label{sec:physical_conditions}

\begin{figure}
    %\centering
    \includegraphics[width=\linewidth]{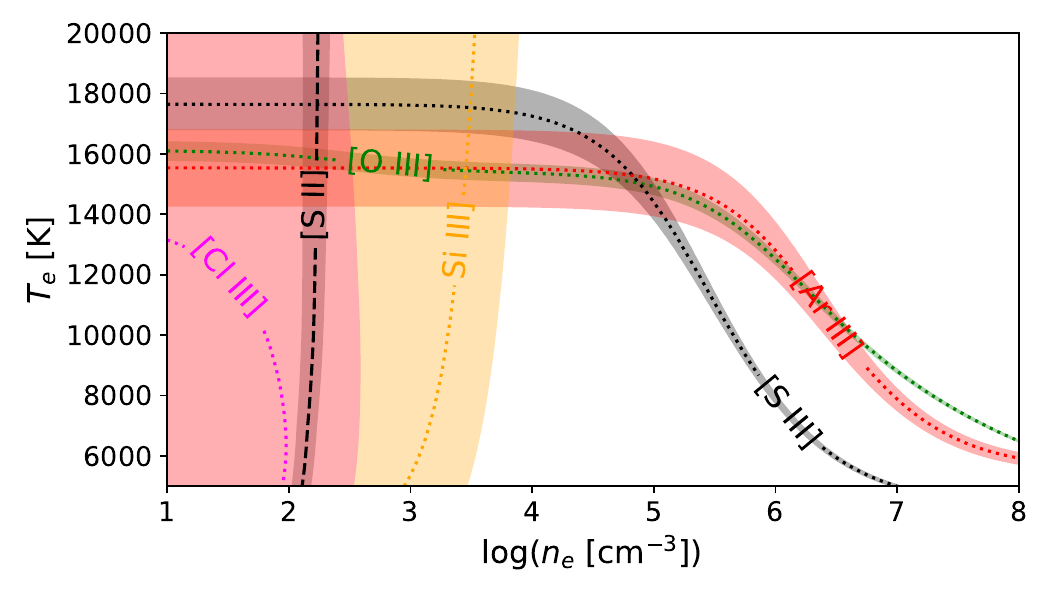}
    \caption{Ionic temperature ($T_e$) and density ($n_e$) diagnostics of UM 461 R1 from optical and UV emission line ratios from S (black contours), Cl (magenta contour), Si (yellow contour), Ar (red contour) and O (green contour). Considering the [S~\textsc{ii}] $\lambda$6716/$\lambda$6731 ratio for density and \ariii\ $\lambda$5192/$\lambda$7136 ratio for temperature yields $n_e=170\pm40$ cm$^{-3}$ and $T_e=15500\pm1300$ K. Using the \siii\ $\lambda$6312/$\lambda$9069 ratio for temperature instead yields $n_e=170\pm40$ cm$^{-3}$ and $T_e=17700\pm900$ K.}
    \label{fig:line_diagnostics}
\end{figure}

We simultaneously calculate the electron density ($n_e$), and electron temperature ($T_e$) from the observed sensitive line ratios using the \textsc{getCrossTemDen} method of \textsc{PyNeb} \citep[version 1.1.28;][]{Luridiana2015}. We generate 10000 Monte Carlo iterations with the \textsc{addMonteCarloObs} method of \textsc{PyNeb} from the emission line fluxes, and the uncertainties are propagated to the electron temperatures, densities, chemical abundances and abundance ratios. The references of the atomic data considered in this work are listed in Appendix~\ref{sec:atomic_data}. The available $n_e$ and $T_e$ sensitive line diagnostics of R1 are shown in Figure~\ref{fig:line_diagnostics}. In Table~\ref{tab:ionic_temperatures}, we include the measured values of $n_e$ and $T_e$ for R1. From the [S~\textsc{ii}] $\lambda$6716/$\lambda$6731 ratio (and the \ariii\ $\lambda$5192/$\lambda$7136 ratio for temperature) we establish an electron density of $n_e=170\pm40$ cm$^{-3}$ and temperature of $T_e($\ariii$)=15500\pm1300$ K. The [\cliii] $\lambda$5538/$\lambda$5518 and [\siiii\ $\lambda1883$/\siiii\ $\lambda1892$ ratios take values ($0.7\pm0.1$ and $1.4\pm0.2$, respectively) that within uncertainties correspond to their low-density limits, where the ratios become insensitive to density (see Figure~\ref{fig:line_diagnostics}). The selection of the $T_e$ diagnostic does not change our density estimations. For the rest of this work, we use the $n_e$ estimated from the [S~\textsc{ii}] doublet.

To account for the stratification of $T_e$ of H~\textsc{ii} regions \citep{Garnett1992}, we assume a three-zone model. This model consists of zones of low, intermediate, and high ionization. We consider the O$^+$, N$^+$, and S$^+$ ions for the low-ionization zone. For the intermediate-ionization zone, we consider S$^{2+}$, Ar$^{2+}$ and Cl$^{2+}$. For the high-ionization zone, we consider O$^{2+}$, Cl$^{3+}$ and Ar$^{3+}$.

In the low-ionization zone of R1 and the individual regions, we assume $T_e(\mathrm{low})\approx T_e$([N~\textsc{ii}]). For the intermediate-ionization zone of R1 we assume $T_e(\mathrm{intermediate})\approx T_e$([Ar~\textsc{iii}]) estimated from the \ariii\ $\lambda$5192/$\lambda$7136 ratio. For the individual regions, we assume $T_e(\mathrm{intermediate})\approx T_e$([S~\textsc{iii}]) as the \ariii\ $\lambda$5192 auroral line is only detected in one of them (region 1). We calculate $T_e$([S~\textsc{iii}]) from the \siii\ $\lambda$6312/$\lambda$9069 ratio. In the high-ionization zone, we assume $T_e(\mathrm{high})\approx T_e$([O~\textsc{iii}]) for all objects. 

In the optical data, the [O~\textsc{iii}] $\lambda$4363 and [N \textsc{ii}] $\lambda$5755 auroral lines, usually employed in the determination of $T_e$ of the high and low ionization zones, respectively, are outside the observed spectral range (see Figure~\ref{fig:Muse_um461_r1}). The $T_e$ of the low ionization zone can also be calculated from \oii\ lines. We detect the [O~\textsc{ii}] $\lambda\lambda$7323, 32 auroral lines, however the [O~\textsc{ii}] $\lambda\lambda$3726, 28 nebular doublet is also out of range. Assuming the temperature relations of \citet{MendezDelgado2023} derived from observations of extragalactic H~\textsc{ii} regions, we use $T_e$(\ariii) to estimate $T_e$([O~\textsc{iii}]) and $T_e$([N~\textsc{ii}]) for R1. As for the individual regions, we estimate $T_e$([O~\textsc{iii}]) and $T_e$([N~\textsc{ii}]) from $T_e$(\siii) and the relations of \citet{MendezDelgado2023}.

In the literature, it is common to assume $T_e$([S~\textsc{iii}])$\approx T_e$([Ar~\textsc{iii}]) \citep[e.g.,][]{Garnett1992} thanks to the similar ionization potentials (IP) required to produce S$^{2+}$ and Ar$^{2+}$ (23.34 eV and 27.63 eV, respectively). In R1, we notice $T_e$([S~\textsc{iii}])$>T_e$([Ar~\textsc{iii}]). This may be due to problems with the atomic data or telluric contamination in \siii\ $\lambda 9069$. Testing different combinations of the transition probabilities and collision strengths available in \textsc{PyNeb} for \siii\ does not change this result. In addition, the optical spectra do not cover the \siii\ $\lambda$9531 line, so it is not possible to test the theoretical intensity ratio to assess the degree of telluric contamination in \siii\ $\lambda$9069. In R1, we detect a prominent Paschen jump around 8200 \AA\ (see Figure~\ref{fig:Muse_um461_r1}, lower panel). The profile of this jump due to the contribution of nebular continuum to the total continuum is often found in star-forming galaxies \citep[e.g.,][]{Guseva2006}. We rule out this jump as the cause of the discrepancy, as we fit the local continuum of the lines for flux measurement.

\citet{MendezDelgado2023} provide two sets of relations, one from photoionization models and the other from their extragalactic \hii regions. We use the latter as their observed $T_e$(\siii) are also systematically higher than $T_e$(\ariii) for $T_e>10000$ K and may not all be free of telluric absorptions. \citet{OrteGarcia2026A&A...711A.165O} also defines temperature calibrations from local star-forming regions, but do not include an explicit $T_e$(\ariii) versus $T_e$([N~\textsc{ii}]) relation. Their $T_e$(\ariii) and $T_e$(\siii) versus $T_e$([O~\textsc{iii}]) relations behave similar to those of \citet{MendezDelgado2023}. Nonetheless, $T_e$([N~\textsc{ii}]) from $T_e$(\siii) using the \citet{OrteGarcia2026A&A...711A.165O} relation is systematically higher. The impact of using this relation is commented in Section~\ref{sec:individual_abundances}. It is important to note these empirical relations have systematic uncertainties due to their fitting to the data.

We estimate the ionization parameter, $U$, by employing the calibrations of \citet[][assuming log(P/k)=5]{Kewley2019} for the optical (\siii\ $\lambda9069+$\siii\ $\lambda9531$)/\sii\ $\lambda\lambda6716,31$ ratio (S$_{32}$)\footnote{Assuming \siii\ $\lambda9069+$\siii\ $\lambda9531$ = $3.47\times$\siii\ $\lambda9069$.}; and for the [Ne~\textsc{iii}] 15 $\mu$m/[Ne~\textsc{ii}] 12 $\mu$m ratio (Ne$_{32}$) using the NIR fluxes of \citet{Cormier2015}. We also use \texttt{HII-CHI-mistry-IR}  \citep[version 3.3;][]{PerezDiaz2024} to infer log($U$) from the fluxes of the [S~\textsc{iv}] 10.5 $\mu$m, [Ne~\textsc{ii}] 12.8 $\mu$m, [Ne~\textsc{iii}] 15.6 $\mu$m, [S~\textsc{iii}] 18.7 $\mu$m, and [S~\textsc{iii}] 33.5 $\mu$m NIR lines and the \oiii\ 88 $\mu$m FIR line. Although we observe \ariv\ and \ariii\ lines, the \ariv\ $\lambda$4711 line required to calculate log($U$) using from the calibrations of \citet{Berg2019} is blended with \hei\ $\lambda$4713. The resulting log($U$) values are listed in Table~\ref{tab:ionic_temperatures}. When inferred from the MIR lines, the ionization parameter is higher (within 3$\sigma$) than when estimated from S$_{32}$. This ratio may indicate that \sii\ comes from a less ionized region due to the low ionization potential needed to create S$^+$ (10.36 eV). However, \citet{Kewley2019} also mentions this ratio should be used with caution at low metallicities due to models underestimating the \sii\ lines.

\subsection{Abundances}
\label{sec:abun}
\begin{deluxetable}{lcc}[h]
%\digitalasset
\tablewidth{0pt}
\tablecaption{Physical conditions, ionic and total abundances of R1.\label{tab:ionic_temperatures}}
\tablehead{
\colhead{Parameter} & \colhead{R1}
}
\startdata
$n_e([$S~\textsc{ii}$])$ $[$cm$^{-3}]$ & 170 $\pm$ 40 \\
$n_e([$Cl~\textsc{iii}$])$ $[$cm$^{-3}]$ & $<320$  \\
$n_e($Si~\textsc{iii}$])$ $[$cm$^{-3}]$ &  $<10^4$ \\
$T_e([$Ar~\textsc{iii}$])$ $[$10$^3$ K$]$ & 15.5 $\pm$ 1.3  \\
$T_e([$S~\textsc{iii}$])$ $[$10$^3$ K$]$ & 17.7 $\pm$ 0.9 \\
$T_e$(high) $[$10$^3$ K$]$ & 17.2 $\pm$ 1.7 \\
$T_e$(low) $[$10$^3$ K$]$ & 13.8 $\pm$ 0.8 \\  
\midrule
log($U)_{\mathrm{S}_{32}}$ & $-2.94\pm0.01$ \\
log($U)_{\mathrm{Ne}_{32}}$ & $-2.48\pm0.13$ \\
log($U)_\mathrm{HII-CHI-mistry-IR}$ & $-2.35\pm0.23$ \\
\midrule
log(O$^+$/H$^+$) & $-5.20 \pm 0.19$   \\
log(O$^{2+}$/H$^+$) & $-4.31\pm 0.10$  \\
log(N$^+$/H$^+$) & $-6.78\pm 0.06$ \\
log(S$^+$/H$^+$) & $-7.01\pm0.05$ \\
log(S$^{2+}$/H$^+$) & $-6.23\pm 0.09$ \\
log(Ar$^{2+}$/H$^+$) & $-6.75\pm 0.07$  \\
log(Ar$^{3+}$/H$^+$) & $-6.97\pm 0.10$ \\
log(Cl$^{2+}$/H$^+$) &  $-8.10\pm0.08$ \\
log(Cl$^{3+}$/H$^+$) &  $-8.12\pm 0.09$ \\
\midrule
log(ICF(N$^+$)) & $0.95\pm0.12$ \\
log(ICF(S$^+$ + S$^{2+}$)) & 0.15 $\pm$ 0.03 \\
log(ICF(Ar$^{2+}$ + Ar$^{3+}$)) & 0.0011 $\pm$ 0.0006 \\
\midrule
12 + log(O/H) & 7.75 $\pm$ 0.11 \\
12 + log(N/H) & 6.17 $\pm$ 0.13 \\
12 + log(S/H) & 5.99 $\pm$ 0.09 \\
12 + log(Ar/H) & 5.46 $\pm$ 0.08 \\
12 + log(Cl/H) & 4.19 $\pm$ 0.09 \\
\midrule
log(N/O) & $-1.57\ \pm$ 0.15 \\
log(S/O) & $-1.77\ \pm$ 0.05 \\
log(Ar/O) & $-2.29\ \pm$ 0.03 \\
log(Cl/O) & $-3.55\ \pm$ 0.04 \\
\midrule
C$^{2+}$/O$^{2+}$  & $0.17 \pm 0.02$  \\
C$^{3+}$/O$^{2+}$  & $0.023 \pm 0.002$ \\
log(C/O)$_{\mathrm{direct}}$ & $-0.73 \pm 0.04$ \\
log(C/O)$_{\mathrm{empirical}}$ & $-0.72 \pm 0.02$ \\
\enddata
\end{deluxetable}
We calculate with \textsc{PyNeb} the ionic abundances (O$^+/$H$^+$, O$^{2+}/$H$^+$, N$^+/$H$^+$, S$^+/$H$^+$, S$^{2+}/$H$^+$, Cl$^{2+}/$H$^+$ and Cl$^{3+}/$H$^+$, Ar$^{2+}/$H$^+$ and Ar$^{3+}/$H$^+$). We assume the ionization correction factors (ICFs) compiled in \citet{PerezMontero2017}, except where noted. These ICFs were calculated for \hii regions and star-forming galaxies. In the following, we summarize the approach taken for deriving the ionic and total elemental abundances of the COS-matched aperture R1 and the individual regions (except where noted).

\uline{Oxygen:} We estimate O$^+/$H$^+$ from the [O~\textsc{ii}] $\lambda\lambda$7323, 32 lines assuming the low-ionization zone temperature $T_e(\mathrm{low})$. O$^{++}/$H$^+$ is estimated from the \oiii\ $\lambda\lambda 4959,5007$ doublet and the high-ionization zone temperature $T_e(\mathrm{high})$. We calculate the total oxygen abundance by assuming
\begin{equation}
    \frac{\mathrm{O}}{\mathrm{H}} = \frac{\mathrm{O}^+ + \mathrm{O}^{2+}}{\mathrm{H}^+}.
    \label{eq:OH}
\end{equation}
The oxygen gas-phase abundance of R1 ($12+\log($O/H$)=7.75\pm0.11$) agrees within uncertainties with the values reported in the literature for the galaxy UM 461 using direct method \citep[$12+\log($O/H$)=7.81\pm0.02$]{Izotov2006} and diagnostics from IR lines \citep[$12+\log($O/H$)=7.92\pm0.25$]{PerezDiaz2024}. 

\uline{Nitrogen:} Assuming $T_e(\mathrm{low})$, the \nii\ $\lambda\lambda 6548,84$ doublet is used to estimate N$^+/$H$^+$. The N$^+$ and O$^+$ ions share similar limits of ionization (14.5 to 29.6 eV and 13.6 to 35.1 eV, respectively). Assuming that N/O $\approx$ N$^+$/O$^+$ \citep{Peimbert1967}, we calculate the total abundance of N/H using the following ICF:
\begin{equation}
    \mathrm{ICF}(\mathrm{N}^+) = \frac{\mathrm{O}}{\mathrm{O}^+}.
    \label{eq:N_ICF}
\end{equation}

\uline{Sulfur:} The S$^+/$H$^+$ ionic abundance is calculated from the \sii\ $\lambda\lambda 6716,31$ doublet and the low-ionization zone temperature $T_e(\mathrm{low})$. To calculate S$^{2+}/$H$^+$, we consider the \siii\ $\lambda 6312$ and \siii\ $\lambda 9069$ lines. For R1 we assume $T_e(\mathrm{intermediate})\approx T_e$([Ar~\textsc{iii}]), and for the individual regions $T_e(\mathrm{intermediate})\approx T_e$([S~\textsc{iii}]) (see Section~\ref{sec:physical_conditions}). We establish the total abundance from S$^+$ and S$^{++}$ using the ICF of \citet{Dors2016}:
\begin{equation}
    \mathrm{ICF}(\mathrm{S}^+ +\mathrm{S}^{2+}) = \left[1-\left(\frac{\mathrm{O}^{2+}}{\mathrm{O}^+ +\mathrm{O}^{{2+}}}\right)^\alpha \right]^{-1/\alpha},
    \label{eq:Stasinska_S_ICF}
\end{equation}
considering $\alpha\approx 3.27$. \citet{PerezDiaz2024} found $12+\log($S/H) = 6.19 $\pm$ 0.38 using IR lines, agreeing with our estimate of $12+\log($S/H$) = 5.99 \pm 0.09$. 

\uline{Argon:} 
The Ar$^{2+}$ ionic abundance is calculated considering the intermediate-ionization zone temperature $T_e(\mathrm{intermediate})$, and the \ariii\ $\lambda 7136$ and \ariii\ $\lambda 7751$ emission lines. For R1 and region 1, we also consider the \ariii\ $\lambda 5192$ auroral line. In the case of Ar$^{3+}$, we only consider the \ariv\ $\lambda 4740$ line, due to the blending of \ariv\ $\lambda 4711$ with \hei\ $\lambda$4713. For Ar$^{3+}$, we assume the high-ionization zone temperature $T_e(\mathrm{high})$. To estimate the total abundance we use the ICF of \citet{PerezMontero2007}:
\begin{equation}
    \mathrm{ICF}(\mathrm{Ar}^{2+} +\mathrm{Ar}^{3+}) = 0.928+0.364(1-x)+\frac{0.006}{1-x},
    \label{eq:PerezMontero_Ar2Ar3_ICF}
\end{equation}
where $x=$ O$^{2+}/($O$^+ +$O$^{2+})$.

We do not detect \ariv\ $\lambda$4740 in three of the individual regions (4, 6 and 7). For these, we first estimate the ionic Ar$^{2+}$ from \ariii\ $\lambda 7136$ and \ariii\ $\lambda 7751$, and then derive the total abundance of Ar by applying the ICF obtained using the following Equation 17 of \citet{PerezMontero2007}:
\begin{equation}
    \mathrm{ICF(Ar}^{2+}) = 0.596 + 0.967(1-x) + \frac{0.077}{1-x},
    \label{eq:icf_ar3}
\end{equation}
where $x=$ O$^{2+}/($O$^+ +$O$^{2+})$.

\uline{Chlorine:} For R1, the ionic abundance of Cl$^{2+}$ is calculated using the [\cliii]\ $\lambda 5518,38$ doublet and the intermediate-ionization zone temperature $T_e(\mathrm{intermediate})$. For Cl$^{3+}$ we assume the high-ionization zone temperature $T_e(\mathrm{high})$ and the \cliv\ $\lambda\lambda 7531,8046$ doublet. A significant fraction of Cl$^+$ is not likely to be present in highly ionized objects \citep[O$^{2+}/$O$\gtrsim0.8$,][]{Orte-Garcia2025}. In the case of R1, O$^{2+}/$O$=0.88\pm 0.03$, thus we calculate the total Cl abundance as
\begin{equation}
    \frac{\mathrm{Cl}}{\mathrm{H}} = \frac{\mathrm{Cl}^{2+} + \mathrm{Cl}^{3+}}{\mathrm{H}^+}.
    \label{eq:Cl}
\end{equation}

In the individual regions, [\cliii] $\lambda 5518$ is only detected in regions 1 and 2, meanwhile the [\cliii] $\lambda5518$ line is detected in regions 1, 2, 3 and 5. The [\civ] lines are only detected in region 1, thus we use Eq.~\ref{eq:Cl} for its total Cl abundance. {\citet{PerezMontero2017} does not include ICFs for chlorine, instead, for} regions 2, 3 and 5 we use the ICF(Cl$^{2+}$) of \citet{Izotov2006} for intermediate $Z$ ($7.2<12+\log$(O/H$)<8.2$) to derive the total chlorine abundance:
\begin{equation}
    \mathrm{ICF(Cl}^{2+}) = 0.814v+0.620+0.131/v,
    \label{eq:izotov_icf_cl}
\end{equation}
where $v=~$O$^+/($O$^+ +$O$^{2+})$.

The resulting ionic and total abundances are shown in Table~\ref{tab:ionic_temperatures}.

\subsection{Abundance ratios\label{sec:abundance_ratios}}

C$^{2+}/$O$^{2+}$ and C$^{3+}/$O$^{2+}$ ratios of R1 were calculated with the dereddened \ciiisemi\ $\lambda\lambda 1908/$\oiiiuv\ $\lambda\lambda1660,6$ and \civ\ $\lambda\lambda1548,50/$\oiiiuv\ $\lambda\lambda1660,6$ line ratios and the high-ionization zone temperature $T_e(\mathrm{high})$ assuming the relations of \citet{PerezMontero2017b}. We estimate direct method C/O by assuming $\frac{\mathrm{C}}{\mathrm{O}} = \frac{\mathrm{C}^{2+}+\mathrm{C}^{3+}}{\mathrm{O}^{2+}}$. We also calculate C/O using the empirical calibration provided in Eq. 7 of \citet{PerezMontero2017b}. The direct and empirical methods yield similar estimates (Table~\ref{tab:ionic_temperatures}).

For estimating the relative abundance ratios of nitrogen, sulfur, argon and chlorine with respect to that of oxygen, we combined the abundance of each of these elements with respect to the oxygen abundances estimated in the Section \ref{sec:abun}. The relative abundance ratios are shown in Table \ref{tab:ionic_temperatures}. In general, our N/O, S/O, Ar/O, and Cl/O abundance ratios of R1 are consistent with the values found in the literature for UM 461. The $\log($N/O$)=-1.56\pm0.15$ we calculate agrees with that reported by \citet[][log(N/O)$=-1.50 \pm 0.06$]{Izotov1999}, and the average values of low metallicity H~\textsc{ii} regions \citep[e.g.,][]{Izotov1999, Kumari2018}. Our log(Ar/O) estimate agrees with the ratio reported by $=-2.27 \pm 0.08$ reported by \citet{Izotov1999}. Also, our log(Cl/O) agrees with \citet{Izotov2006}, who reported log(Cl/O)$=-3.65 \pm 0.14$ in UM 461. 

The observed $\log($S/O$)=-1.77\pm0.05$ is lower than the solar value of $-1.57 \pm 0.05$ \citep{Asplund2021}. Nonetheless, our estimate is consistent with the mean values of 441 star-forming galaxies (log(S/O)$=-1.63 \pm 0.13$) and 492 H~\textsc{ii} regions (log(S/O)$=-1.54 \pm 0.19$) found by the DEep Spectra of Ionised REgions DAtabase Extended project \citep[DESIRED-E;][]{Esteban2025}. Our log(S/O) estimate is systematically lower than the UM 461 value from \citet{Izotov1999}. This will be discussed in Section~\ref{sec:individual_abundances}. 

The projected radial distribution of oxygen abundance and the N/O, S/O, and Ar/O abundance ratios of the seven individual regions will also be discussed in Section~\ref{sec:individual_abundances}.

\subsection{Emission-line diagnostic diagrams \label{sec:diagnostic_diagrams}}

\begin{figure}
    \centering
    \includegraphics[width=\linewidth]{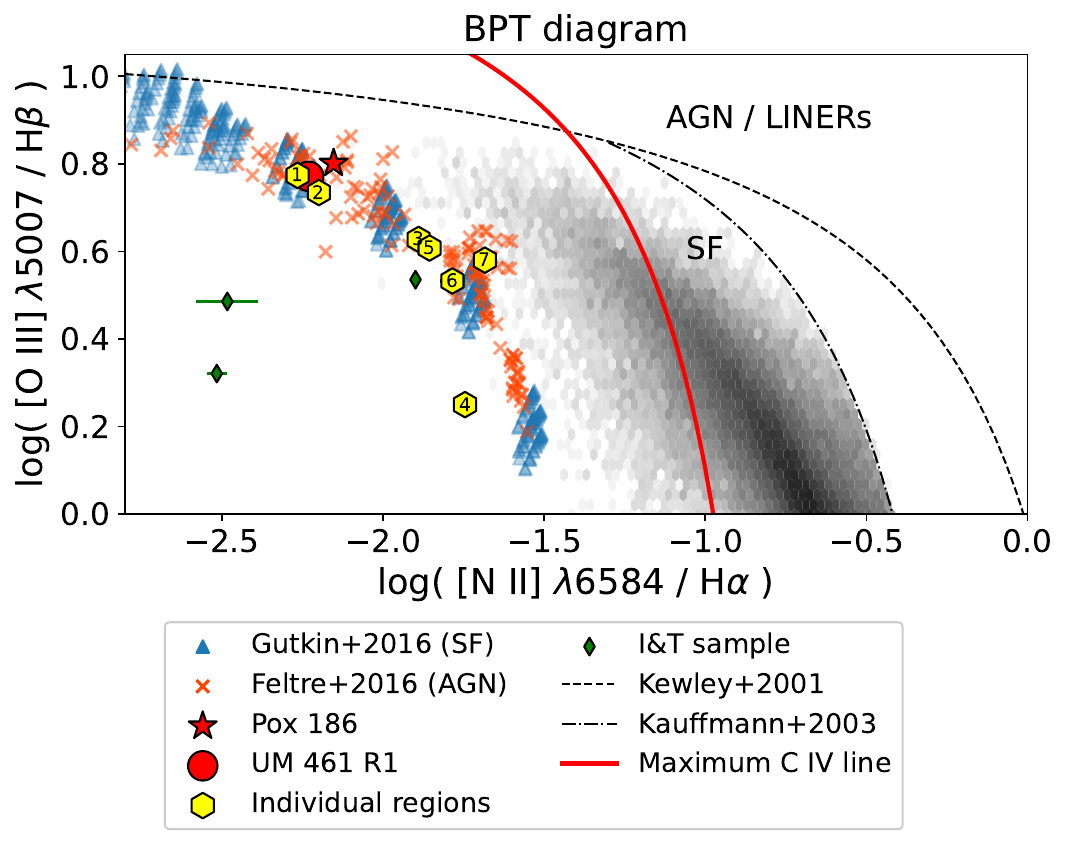}
    \caption{BPT \citep{Baldwin1981} diagnostic diagram. The COS-matched aperture R1 (red circle) and the individual regions (yellow hexagons), and the I\&T sample \citep{Izotov1998,Izotov1999} of supergiant H~\textsc{ii} regions with observed \heii\ $\lambda$4686, \ciiisemi\ $\lambda\lambda1908$ and \siiii\ $\lambda\lambda 1888$ (green diamonds) are presented. $Z=0.002$ photoionization models of star-forming regions (blue triangles) from \citet{Gutkin2016} and AGNs (orange crosses) from \citet{Feltre2016} are presented in the background. Lines of maximum star-formation of \citet[][dashed line]{Kewley2001} and \citet[][dashed-dotted line]{Kauffmann2003} are also shown. The upper limit of the region in which the nebular component of the \civ\ $\lambda\lambda 1548,50$ doublet is expected in star-forming galaxies is shown as a red solid line (see Section~\ref{sec:civ_profile}). The distribution of star-forming galaxies from SDSS-DR8 \citep{Aihara2011} is shown in the background.} 
    \label{fig:BPT}
\end{figure}

In this section, we analyze the main source of ionizing radiation of R1 and the individual regions within UM 461 by employing diagnostic diagrams involving emission line ratios and equivalent widths. We also consider literature data of star-forming galaxies exhibiting similar optical+UV properties for comparison in these diagrams. In particular, we use Pox 186, the pilot galaxy in our multi-wavelength study of local EoR analogues, presented in \citet{Kumari2024}. We include regions from \citet{Izotov1998,Izotov1999}, the largest known compilation of BCD H~\textsc{ii} regions hosting intense bursts of star-formation, low metallicities and with available optical and UV data. From this sample, we select the regions with measurements of the \heii\ $\lambda 4686$, \ciiisemi\ $\lambda\lambda1908$ and \siiii\ $\lambda\lambda 1888$ lines. Three of their objects (I\&T sample hereafter) include these lines: SBS 1415+437, SBS 0335-052, and the northwest component of I Zw 18 (I Zw 18 NW hereafter). The UV data of the I\&T sample come from \citet{Izotov1999}. Optical data of SBS 1415+437 and SBS 0335-052 are taken from \citet{Izotov1998}. We use the optical data of \citet{Kehrig2016} for I Zw 18 NW, as the \nii\ $\lambda6584$ line is missing from \citet{Izotov1999}.

We perform a detailed comparison of these datasets with the photoionization models from \citet{Gutkin2016} and \citet{Feltre2016}, to characterize the properties of the ionizing source. \citet[][G16 hereafter]{Gutkin2016} calculated a grid\footnote{\url{https://www.iap.fr/neogal/sf-models.html}.} of photoionization models of star-forming galaxies with version c13.03 of \texttt{CLOUDY} \citep[last described by][]{Ferland2013}, using population synthesis models from an updated version of the \citet{Bruzual2003} models as ionizing sources. The G16 models cover ionization parameters $-1 \leq\log U_\mathrm{S} \leq -4$ at the Str{\"o}mgren radius\footnote{The ionization parameter at different radii is the same irrespective of the geometry considered, plane-parallel versus spherical \citep{Kewley2019}.}, dust-to-metal mass ratios 0.1 $\leq\xi_\mathrm{d}\leq$ 0.5, densities $10\leq n_\mathrm{H}\leq10^4$ cm$^{-3}$, C/O ratios 0.10 $\leq (\mathrm{C}/\mathrm{O})/(\mathrm{C}/\mathrm{O})_\odot$\footnote{G16 consider (C/O)$_\odot=0.44$.} $\leq$ 1.40 and upper limits of the initial stellar mass function of $m_\mathrm{up}=100-300\,\mathrm{M}_\odot$. The active galactic nuclei (AGN) narrow-line region models of \citet[][F16 hereafter]{Feltre2016}\footnote{\url{https://www.iap.fr/neogal/agn-models.html}} were calculated with the same version of \texttt{CLOUDY} as G16, and approximating the emission from the accretion disc of an AGN by a broken power law. The F16 models cover power-law indices of the ionizing spectrum $-1.2\leq \alpha \leq -2.0$, densities $10^2\leq n_\mathrm{H}\leq10^4$ cm$^{-3}$, and the same ionization parameters and dust-to-metal mass ratios as G16.  In this work, we only consider the G16 and F16 models with total metallicity of $Z_\mathrm{ISM}=0.002$ and $\xi_\mathrm{d}=0.1,~0.3$ (SF models and AGN models, hereafter). These models cover the gas-phase oxygen abundances of $7.83\leq$ 12 + log(O/H)$_\mathrm{gas} \leq 7.91$ (depending on depletion), similar to those observed in UM 461 and Pox 186. The dust-to-metal mass ratios are expected to decrease with metallicity, with $\xi_\mathrm{d,\odot}\sim0.4$ \citep{Hammanowicz2024,Isobe2026}. Thus, we do not consider the models with $\xi_\mathrm{d}=0.5$. 

\begin{figure}
    \centering
    \includegraphics[width=\linewidth]{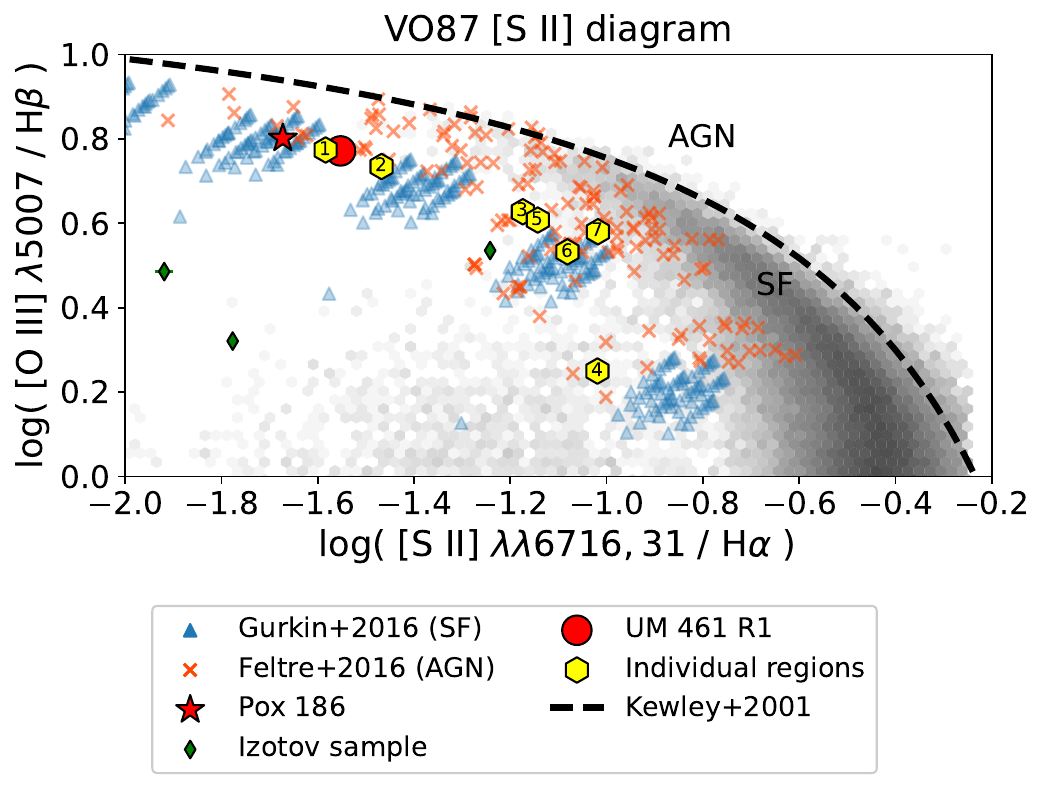}
    \caption{VO87 [S~\textsc{ii}] \citep{Veilleux1987} diagnostic diagram. We include the location of UM 461 R1 (red circle), the individual regions (yellow hexagons), Pox 186 (red star) and the I\&T sample (green diamonds). In the background, we plot the distribution of star-forming galaxcies of SDSS-DR8 \citep{Aihara2011}, the $Z=0.002$ models of star-forming regions (blue triangles) from \citet{Gutkin2016} and AGNs (orange crosses) of \citet{Feltre2016}. The black dash-dotted line represents the maximum star-formation line of \citet{Kewley2001}.}
    \label{fig:VO87}
\end{figure}

In Figures~\ref{fig:BPT} and \ref{fig:VO87}, we show R1, Pox 186 and the individual H~\textsc{ii} regions of UM 461 in the BPT \citep{Baldwin1981} and VO87 \sii\ \citep{Veilleux1987} diagnostic diagrams, respectively. We show the maximum star-formation lines of \citet[][only in Figure~\ref{fig:BPT}]{Kauffmann2003} and \citet[][in both figures]{Kewley2001}. These lines are commonly used to distinguish between star formation (SF) and the presence of an AGN as the main source of ionizing photons in unresolved galaxies. Both theoretical and empirical maximum starburst lines suggest that the individual regions within UM 461, including COS-matched region (R1) exhibit the emission line ratios dominated by SF. We include the SF and AGN models in the optical emission line ratio diagnostic diagrams (Figs.~\ref{fig:BPT} and \ref{fig:VO87}). It is evident that both SF and AGN models predict similar line ratios, as these optical diagrams do not work as intended on low metallicity regions ($Z\lesssim0.008$; F16).

\begin{figure*}
    \centering
    \includegraphics[width=\linewidth]{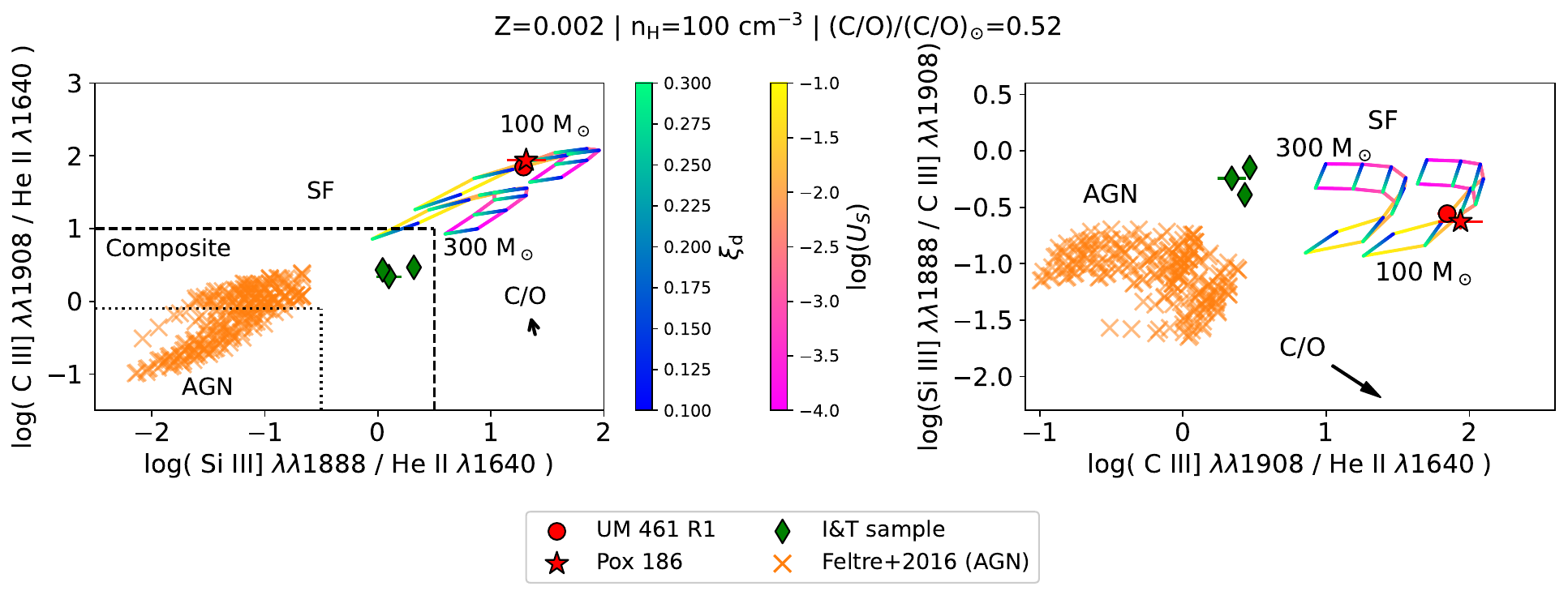}
    \caption{UV diagnostic diagrams. \textit{Left panel:} \ciiisemi/\heii\ versus \siiii/\heii\ diagnostic from \citet{Hirshcmann2019}. \textit{Right panel:} \siiii/\ciiisemi\ versus \ciiisemi/\heii\ diagnostic. $Z=0.002$ AGN models from \citet{Feltre2016} shown as orange crosses. Models of star-forming regions from \citet{Gutkin2016} with upper mass limits of 100 and 300 $M_\odot$ are shown as grids of ionization parameter (log($U_S$)) versus depletion ($\xi_\mathrm{d}$). The fixed values of the displayed \citet{Gutkin2016} models are shown in the title. Arrows show the direction of increasing C/O in the SF models.}
    \label{fig:uv_diagrams}
\end{figure*}

This overlap of SF and AGN models in optical diagnostic diagrams prevents us from identifying the nature of the ionizing source in UM 461 and Pox 186. UV diagnostic diagrams have been proposed in the literature as an alternative to solve this issue (e.g., F16). Many of these diagrams use the He~\textsc{ii} $\lambda$1640 recombination line instead of optical Balmer H lines. He~\textsc{ii} $\lambda$1640 is not visible in our COS G160M spectrum (see Figure~\ref{fig:COS_G160M_um461}). In principle, this line is stronger than the optical He~\textsc{ii} $\lambda$4686, however, it remains hidden due to the strength of the stellar continuum in the UV regime of star-forming regions. Nonetheless, we can estimate the He~\textsc{ii} $\lambda$1640 flux of R1 from the observed optical He~\textsc{ii} $\lambda$4686 line using \textsc{PyNeb}. Assuming Case B recombination for an electronic temperature of $T_e=$ 15000 K and density of 100 cm$^{-3}$, the He~\textsc{ii} $\lambda$1640/4686 ratio takes a value of 6.96. In the case of Pox 186, we use the \heii\ $\lambda1640$ flux from \citet{Rogers2023}. In Figure~\ref{fig:uv_diagrams} we plot the \ciiisemi\ $\lambda\lambda 1908/$\heii\ $\lambda 1640$ versus \siiii\ $\lambda\lambda1888/$\heii\ $\lambda 1640$ diagnostic from \citet{Hirshcmann2019}, as well as in the \siiii\ $\lambda\lambda1888/$\ciiisemi\ $\lambda\lambda 1908$ versus \ciiisemi\ $\lambda\lambda 1908/$\heii\ $\lambda 1640$ diagnostic. In both UV diagnostics, the SF and AGN models are clearly separated from each other. In contrast to the optical diagrams (Figures \ref{fig:BPT} and \ref{fig:VO87}), the line ratios for UM 461 R1 and Pox 186 are reproduced only by the SF models. This confirms that both UM 461 and Pox 186 are ionized by star formation instead of AGN activity.

\begin{figure}
    \centering
    \includegraphics[width=\linewidth]{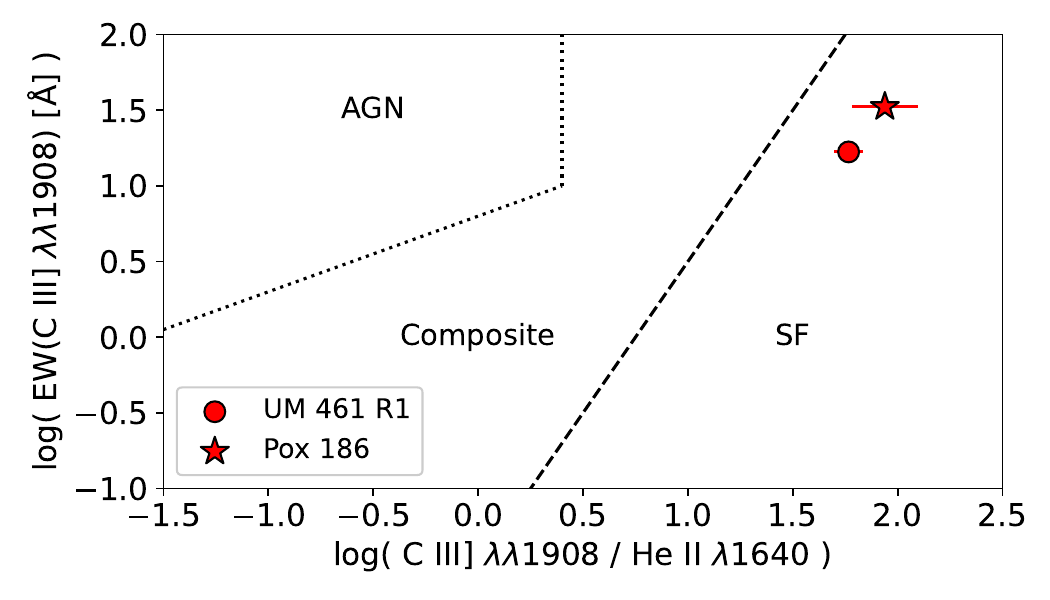}
    \caption{EW(\ciiisemi) versus \ciiisemi/\heii\ UV diagnostic diagram and lines from \citet{Hirshcmann2019}. UM 461 R1 and Pox 186 lie in the region of galaxies dominated by star-formation.}
    \label{fig:H19_EW-C3}
\end{figure}

These UV diagrams are more useful than the traditional optical diagrams to distinguish SF-dominated from AGN galaxies at low metallicities. However, there are caveats to consider when using these \ciiisemi\ and \siiii\ lines. First, both lines of the \siiii\ $\lambda\lambda1888$ doublet are not always detected simultaneously. In the I\&T sample \siiii\ $\lambda1892$ is only detected in SBS 0335-052. Fortunately, the ratio of emissivities of these lines is $\sim 0.55$ for densities $\lesssim 10^3$ cm$^{-3}$ (Figure~\ref{fig:Si3_doublet_ratio}). \citet{Izotov1999} adopt F($\lambda$1883 + 1892) = 1.67 F($\lambda 1883$) for SBS 415+437 and I Zw 18 NW. The second caveat is the total Si/C abundance ratio. In G16, all metals except for N and C are scaled linearly with metallicity. C/O takes one of nine logarithmically spaced values from 0.10 to 1.40 times their solar value of (C/O)$_\odot = 0.44$. To conserve the total metallicities, all other metals (including Si) have their abundances slightly reduced. The effect of increasing C/O in the models is an increase in the \ciiisemi\ lines followed by a slight reduction in \siiii\ lines, depicted by the arrows in Figure~\ref{fig:uv_diagrams}. Meanwhile, the third consideration is that both Si and C can be depleted onto dust. The relative degree of C depletion compared to Si is uncertain \citep[see Figure 5 of ][]{Jenkins2009}. In G16 and F16, the amount of metals depleted is varied by the $\xi_\mathrm{d}$ parameter and Si has a higher degree of depletion than C, thus increasing $\xi_\mathrm{d}$ decreases the \siiii/\heii\ and \siiii/\ciiisemi\ ratios more than the \ciiisemi/\heii\ ratio does. 

The nature of the ionizing sources is also investigated via the equivalent widths of different emission lines \citep[e.g.,][]{Hirshcmann2019}. In Figure~\ref{fig:H19_EW-C3} we show EW(\ciiisemi) versus \ciiisemi/\siiii\ UV diagram as proposed by \citet{Hirshcmann2019} to distinguish between SF, composite, and AGN, though the equivalent widths of lines for the SF and AGN models are not publicly available. On this diagram, the dashed and dotted lines separate different ionizing sources. Figure~\ref{fig:H19_EW-C3} also indicates that R1 and Pox 186 are ionized by stars, in agreement with the previous UV emission line ratios diagrams described earlier.

\subsection{C~\textsc{iv} line profile and nebular emission \label{sec:civ_profile}}

\begin{figure}
    \centering
    \includegraphics[width=\linewidth]{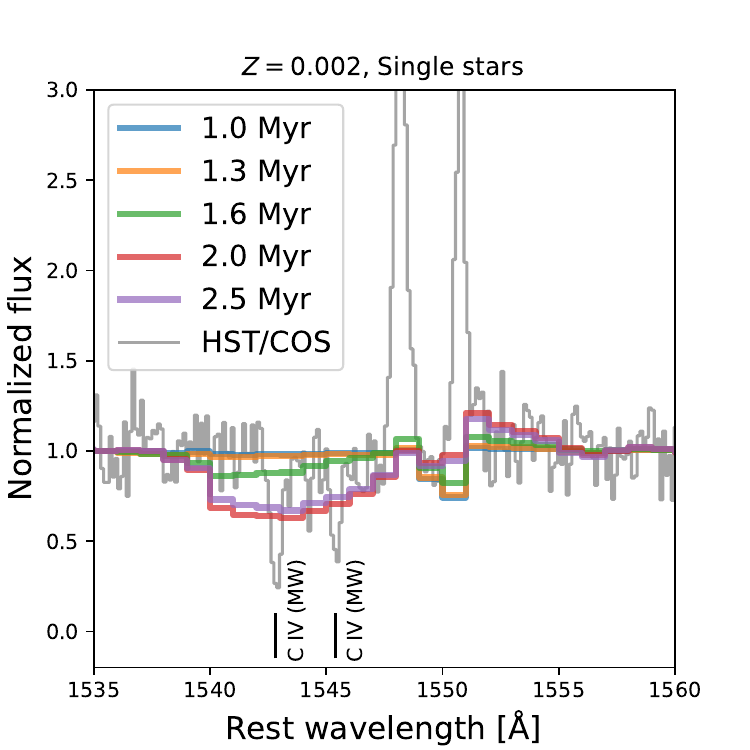}
    \caption{Comparison of the observed \civ\ profile with BPASS models of single stars of $Z=0.002$ at ages of 1 (blue), 1.3 (orange), 1.6 (green), 2 (red), and 2.5 Myr (purple). We indicate the location of the Galactic \civ\ absorption lines. \textit{HST}/COS spectrum is rebinned by 10 pixels for better visibility.}
    \label{fig:civ_profile}
\end{figure}

In Figure~\ref{fig:civ_profile}, we compare the observed \civ\ profile with the stellar population synthesis models of BPASS v2.2.1 \citep{Stanway2018} with single stars of $Z=0.002$ metallicity and an upper mass-limit of 300 $M_\odot$. The absence of a strong P-Cygni profile indicates the presence of a young ($<2$ Myr) stellar population. The nebular \civ\ doublet dominates the line profile.

\begin{figure}
    \centering
    \includegraphics[width=\linewidth]{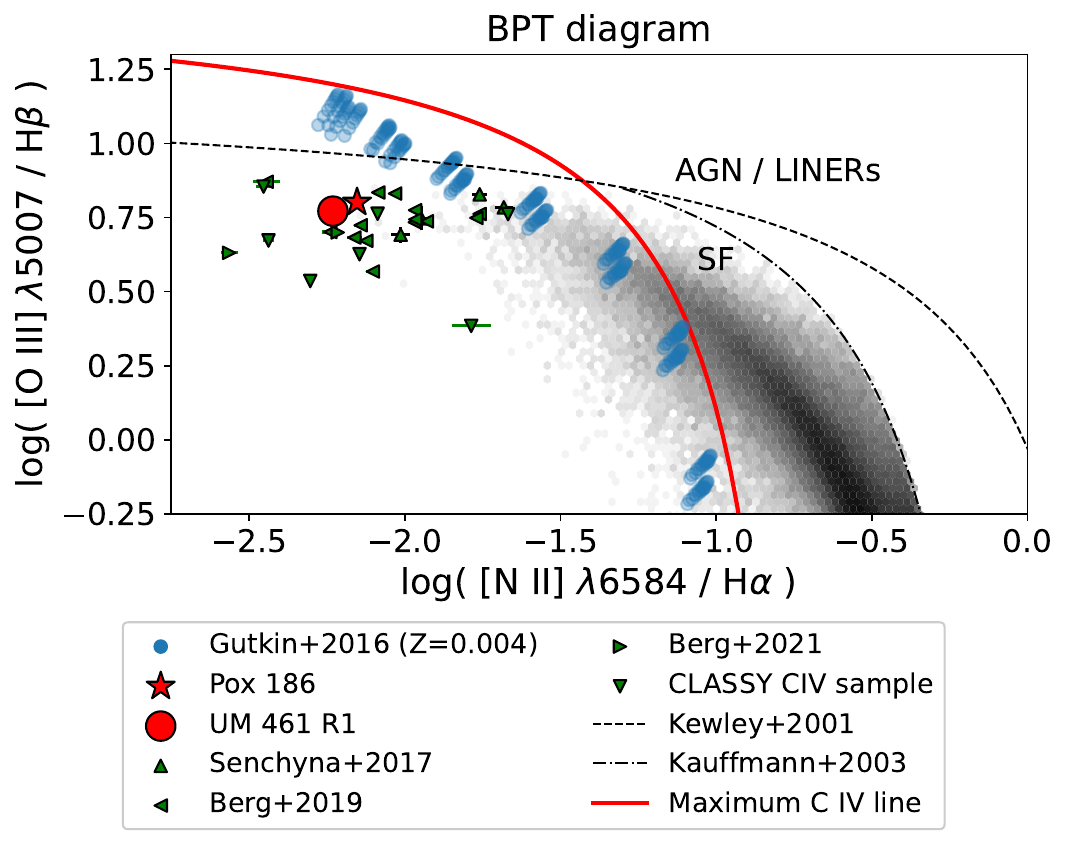}
    \caption{Upper limit of galaxies with nebular \civ\ profiles (red solid line) in the BPT diagram. $Z=0.004$ models of G16 with $\xi_{\rm{d}}=0.3$ used to define this curve are plotted as blue circles. We show the location of UM 461 R1 (red circle) and Pox 186 (red star) in this diagram. In green triangles, the location of star-forming galaxies with nebular \civ\ lines of \citet[][up-pointing]{Senchyna2017}, \citet[][left-pointing]{Berg2019}, \citet[][right-pointing]{Berg2021}, and CLASSY \citep[][down-pointing]{Mingozzi2024}. Distribution of star-forming galaxies from SDSS-DR8, diagnostic lines from \citet{Kewley2001} and \citet{Kauffmann2003} are shown for reference.}
    \label{fig:civ_maximum_limit}
\end{figure}

This nebular component is only observed in ionized regions with oxygen abundances below that of the SMC \citep[$12+\log$(O/H$)\lesssim 8.01$;][]{Senchyna2017,Berg2019,Berg2021,Clark2024}. This allows us to define an upper limit to the area in the BPT diagram where galaxies with potential contribution from the nebular \civ\ doublet are located. The gas-phase O abundance of the $Z=0.004$ models with $\xi_{\rm{d}}=0.3$ ($12+\log$(O/H$)=8.14$) is above that of the current highest-metallicity galaxy with observed nebular \civ\ emission, SB 182 \citep[$12+\log$(O/H$)=8.01$,][]{Senchyna2017}. We fit the envelope of these models to define the upper limit as:
\begin{equation}
    y = \frac{0.51}{x+0.64}+1.52,
    \label{eq:civ_max_line}
\end{equation}
for $x<-0.64$, where $x = \log($\nii\ $\lambda 6584/$H$\alpha$) and $y = \log($\oiii\ $\lambda 5007/$H$\beta$). This equation indicates the upper limit in the BPT diagram where the nebular \civ\ can be detected. This upper limit and the G16 models employed for its definition are plotted in Figure~\ref{fig:civ_maximum_limit}. Along this curve, we plot the galaxies with nebular \civ\ emission from \citet{Senchyna2017}, \citet{Berg2019,Berg2021}, and the Cosmic Origins Spectrograph Legacy Archive Spectroscopic SurveY \citep[CLASSY,][]{Mingozzi2024}. All of these galaxies as well as UM 461 R1 and Pox 186 (which possess nebular \civ) fall below our curve.

\begin{figure}
    \centering
    \includegraphics[width=\linewidth]{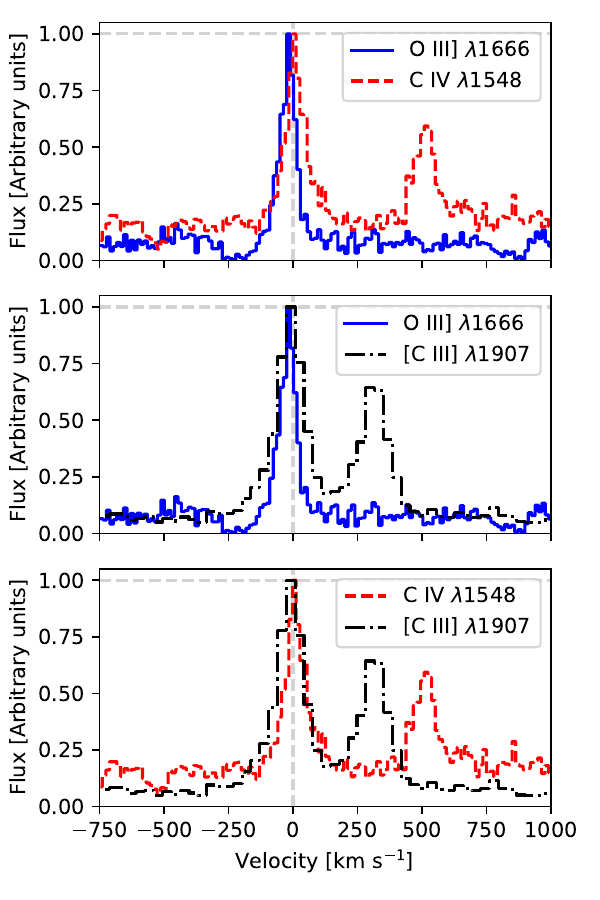}
    \caption{Comparison of velocity profiles relative to \oiiiuv\ $\lambda 1666$ (blue solid curve), \civ\ $\lambda 1548$ (red dashed curve) and [C~\textsc{iii}] $\lambda 1907$ (black dash-dotted curve), normalized by their peak fluxes.}
    \label{fig:UV_line_profiles}
\end{figure}

\begin{figure}
    \centering
    \includegraphics[width=0.95\linewidth]{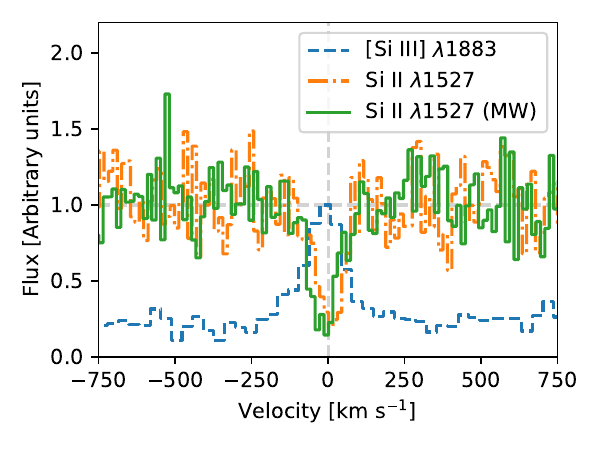}
    \caption{Comparison of velocity profiles of R1 [\siiii\ $\lambda 1883$ emission (dashed blue curve) and Si~\textsc{ii} $\lambda 1527$ absorption (dash-dotted orange curve) and foreground Milky Way Si~\textsc{ii} $\lambda 1527$ absorption (solid green curve). \siiii\ $\lambda 1883$ is normalized by its peak flux. The absorption lines are normalized by the median flux in the velocity range of $-1000$ to 1000 km s$^{-1}$. No signature of outflow is observed for R1.}
    \label{fig:Si_absorption_emission}
\end{figure}

\subsection{Comparison of UV line profiles \label{sec:uv_line_profiles}}

In Figure~\ref{fig:UV_line_profiles}, we compare the normalized velocity profiles of \oiiiuv\ $\lambda 1666$, \civ\ $\lambda 1548$ and [C~\textsc{iii}] $\lambda 1907$ of UM 461 R1. \citet{Kumari2024} report that each component of the \civ\ $\lambda\lambda 1548,50$ doublet has a clear double peak in the HST/COS G165M spectrum of Pox 186. This is not the case in UM 461 R1. The observed peaks of \oiiiuv\ $\lambda 1666$ and \civ\ $\lambda 1548$ are separated by $\sim$10 km s$^{-1}$, however, this is lower than the observed velocity resolution of HST/COS data. In Figure~\ref{fig:Si_absorption_emission}, we compare the velocity profiles of the [\siiii\ $\lambda 1883$ emission and Si~\textsc{ii} $\lambda$1527 absorption lines of R1, and the Milky Way (MW) Si~\textsc{ii} 1527 absorption. The MW absorption lines are expected to be unresolved at the resolution of COS G160M/1623 ($R\sim 14000$), but from our Galactic Si~\textsc{ii} $\lambda$1527 line, we measure $R\sim 2800$, corresponding to a velocity resolution of $\sim 110$ km s$^{-1}$. This reduction in resolution is consistent with observations of unresolved sources \citep[e.g.,][]{Leitherer2013}. Figures~\ref{fig:UV_line_profiles} and \ref{fig:Si_absorption_emission} do not show any signatures of line broadening.

\section{Discussion \label{sec:discussion}}

\subsection{He~\textsc{ii} in UM 461 and lower limit of WN stars \label{sec:heii_discussion}}

The detection of nebular \heii\ $\lambda 4686$ in R1 implies the presence of ionizing photons with E $>$ 54 eV. We analyze the possible sources of this hard ionizing radiation.

\begin{figure}
    \centering
    \includegraphics[trim={0.3cm 0 0 0},clip, width=\linewidth]{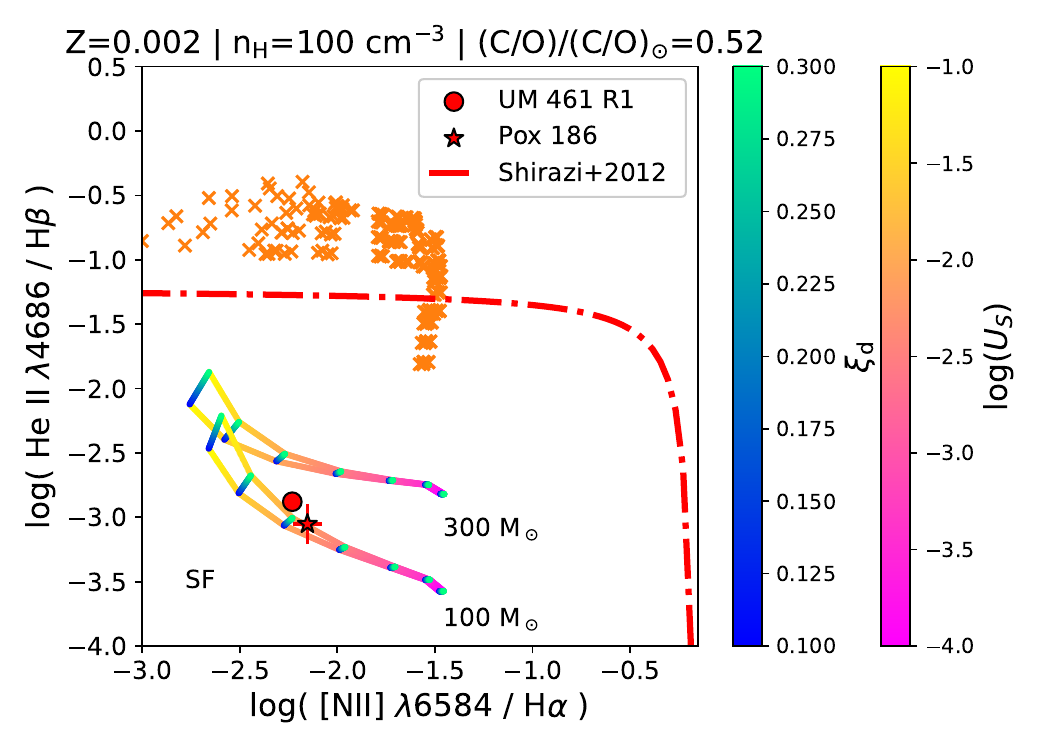}
    \caption{Location of UM 461 R1 (red circle) and Pox 186 (red star) in the He~\textsc{ii}/H$\beta$ versus [N~\textsc{ii}]/H$\alpha$ diagnostic diagram. $\xi_D$ versus ionization parameter ($U_S$) grid of \citet{Gutkin2016} SF models is plotted. The fixed parameters are shown in the title. The $Z=0.002$ AGN models of \citet{Feltre2016} are shown in orange crosses. The SF-AGN separation line of \citet{Shizari2012} is also shown. The uncertainties associated with R1 are smaller than the marker.}
    \label{fig:heiihb_vs_niiha_grid}
\end{figure}

Figure~\ref{fig:heiihb_vs_niiha_grid} shows UM 461 R1 and Pox 186 in the \heii\ $\lambda$4686 /H$\beta$ versus [N\,\textsc{ii}] $\lambda$6584 /H$\alpha$ diagnostic diagram. We convert the \heii\ $\lambda$1640 of Pox 186 to \heii\ $\lambda$4686 using the Case B ratio described in Section \ref{sec:diagnostic_diagrams}. In this plot, we show the same models as in Figure~\ref{fig:uv_diagrams}. We also include the \citet{Shizari2012} SF-AGN separation line. The emission line ratios of Pox 186 are well reproduced by the SF models with upper mass limits of 100 $M_\odot$. Meanwhile, the R1 ratios are not reproduced simultaneously within uncertainties by the models with this upper mass limit. \heii/H$\beta$ at a given \nii/H$\alpha$ increases with metal depletion, implying that a higher depletion could reproduce the R1 observation. However, we expect $\xi_\mathrm{d}$ to decrease from 0.41 at solar metallicity to 0.1 at $Z_\odot/10$ \citep{Hammanowicz2024}. The effect of modifying C/O and gas density is negligible in this diagram, hence the only parameter left is the stellar upper mass limit. The models with stars up to 300 $M_\odot$ produce a harder ionizing field than the 100 $M_\odot$ ones, translating into higher \heii/H$\beta$ ratios. Although these ratios are higher than those of UM 461, this hints that models that include Very Massive Stars (VMS; M$>100\,M_\odot$) with upper masses between 100-300 $M_\odot$ should be able to reproduce this data point. We rule out the possibility of the observed \heii/H$\beta$ ratio being reproduced by using other stellar atmosphere models in G16 models, based on the analysis by \citet{SimonDiaz2008}, which shows that \texttt{WM-Basic} \citep{Pauldrach2001} models used in G16 have the highest or comparable $Q(\rm{He}^1)/Q(\rm{H}^0)$ ratios as other existing models and would thereby result in the highest or comparable \heii/H$\beta$ ratio as another existing atmospheric model.

\citet{Conti1991} classified UM 461 as a Wolf-Rayet (WR) galaxy based on a spectrum of J. Salzer (private communication), claiming detection of a broad He II $\lambda$4686 emission. However, we do not observe the blue WR bump in our optical spectra, in agreement with \citet{Izotov1998}, who did not observe this feature in UM 461 either. Pox 186 was also described as a WR galaxy by \citet{Conti1991} based on the detection of excess emission in the range of 4600-4711 \AA\ by \citet{Kunth1985}. However, \citet{Kumari2024} excludes Pox 186 as a WR galaxy using HST/COS and GMOS-IFU data.

Nonetheless, the observed \heii\ $\lambda$4686 line may also be attributed to weak-winded WNE stars. \citet{GonzalezTora2025} shows that the broad blue WR bump due to one of these stars is completely diluted when using integrated fluxes from regions with a diameter $>24$ pc. The effect can be exacerbated as a result of crowding with other bright stars. Considering the angular scale at the distance of UM 461 is 74 pc arcsec$^{-1}$ (see Section~\ref{sec:distance}), the PSA integrates fluxes over a 185 pc diameter, 7.5 times the diameter needed to resolve the blue bump of weak-winded WNE stars. For comparison, the individual MUSE pixel corresponds to a physical scale of 15 pc at the distance of UM 461.

At the luminosity distance of UM 461, $L_{\mathrm{He\textsc{ii}~\lambda 4686}} = 3.30\times 10^{36}$ erg s$^{-1}$. The \heii\ ionizing flux ($Q_\mathrm{He\textsc{ii}}$) can be estimated from this luminosity using Eq.~1 of \citet{GonzalezTora2025}:
\begin{equation}
    Q_\mathrm{He\textsc{ii}}\approx L_{\mathrm{He\textsc{ii}}~\lambda 4686}/3.66\times 10^{-13}\,\mathrm{erg}.
\end{equation}
We obtain $Q_\mathrm{He\textsc{ii}}=9.00\times10^{48}\mathrm{~s}^{-1}$. Assuming the He~\textsc{ii} ionizing flux of a single WN3 star at SMC metallicity is $\log(Q_\mathrm{He\textsc{ii}}/\mathrm{s}^{-1}) \sim 48.5$ \citep{GonzalezTora2025}, this would yield a lower limit of $\sim 4$ WNE stars in R1.

\subsection{Abundance patterns \label{sec:individual_abundances}}

\begin{figure}
    \centering
    \includegraphics[width=0.45\textwidth]{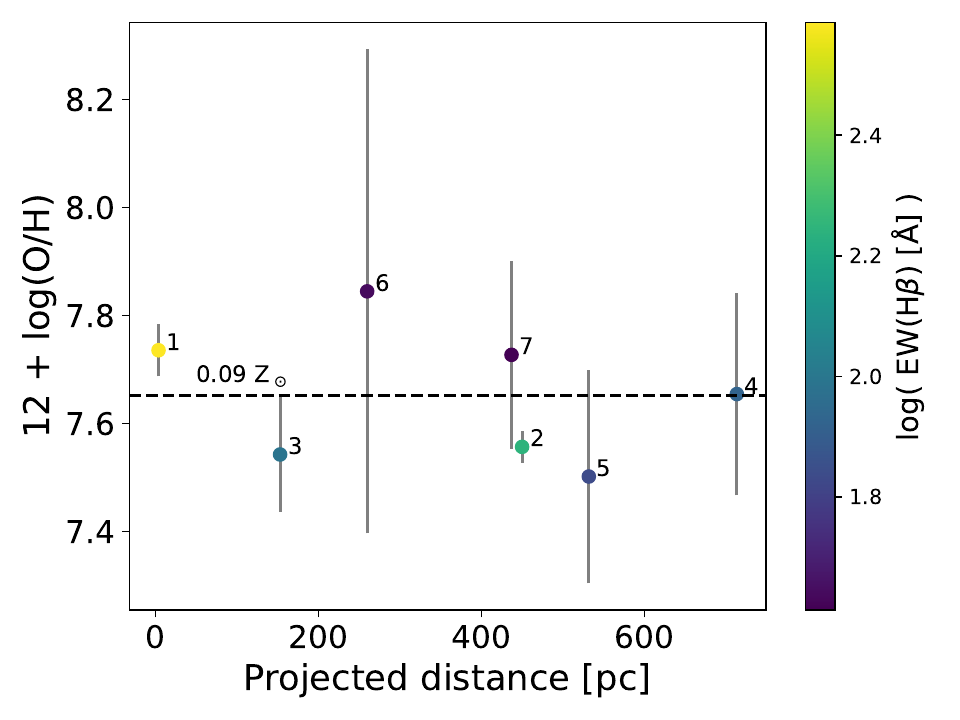}
    \includegraphics[width=0.45\textwidth]{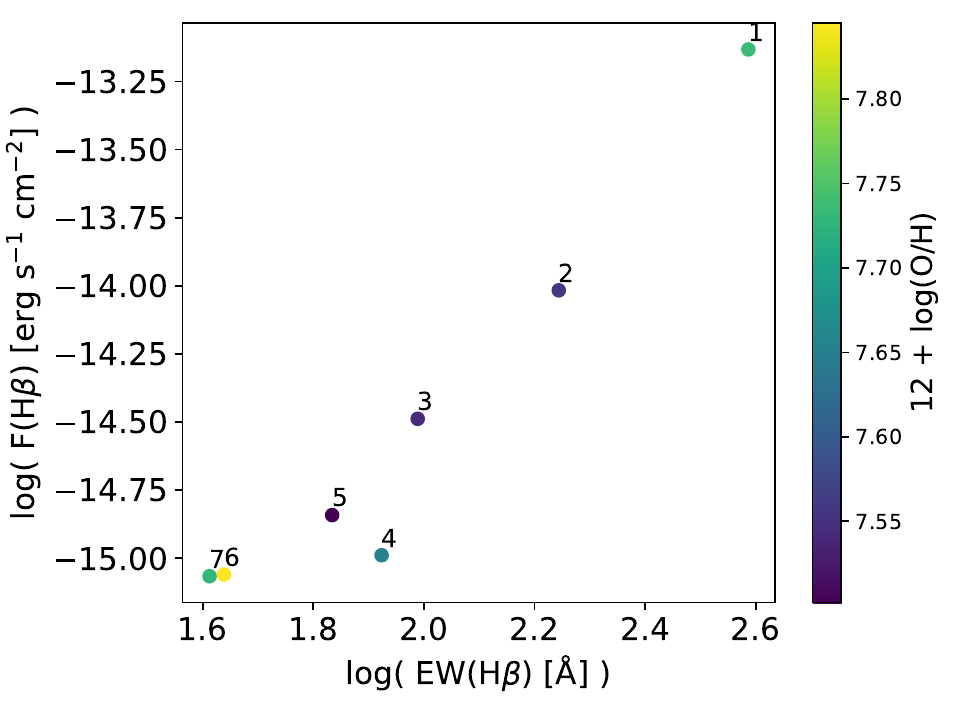}
    \caption{\textit{Top panel:} Projected radial variation of gas-phase metallicities for the seven H~\textsc{ii} regions, where the numerals on each point correspond to those identified in Figure \ref{fig:MUSE_Ha}. The data-point are color-coded with respect to the EW(H$\beta$). The dashed horizontal line indicates the mean gas-phase metallicity of all regions. \textit{Bottom panel:} Variation of equivalent width H$\beta$ with respect to H$\beta$ line flux for the seven regions extracted with \textsc{ifuanal} (Figure \ref{fig:MUSE_Ha}). The regions are color-coded according to their oxygen gas-phase abundance.}
    \label{fig:OH}
\end{figure}

\begin{figure*}
    \centering
    \includegraphics[width=0.45\textwidth]{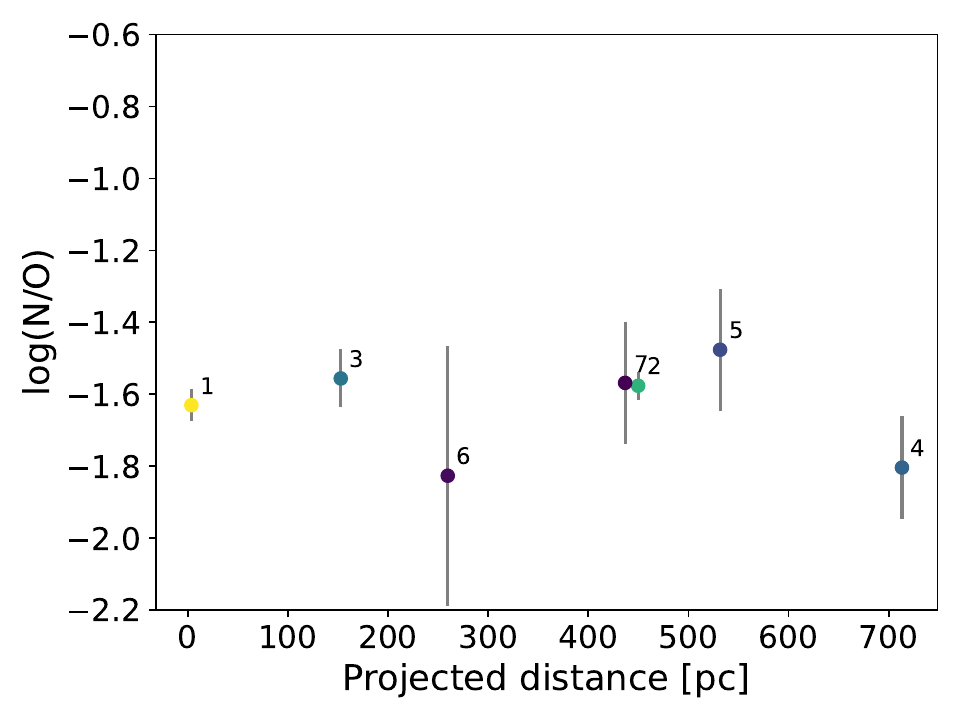}
    \includegraphics[width=0.45\textwidth]{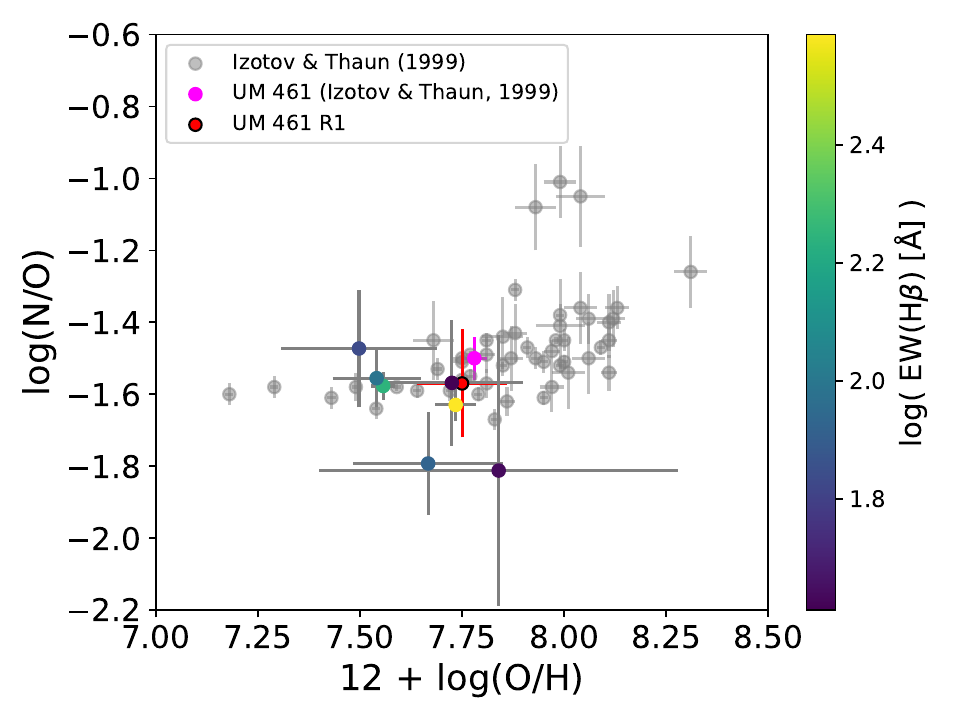}
    
    \includegraphics[width=0.45\textwidth]{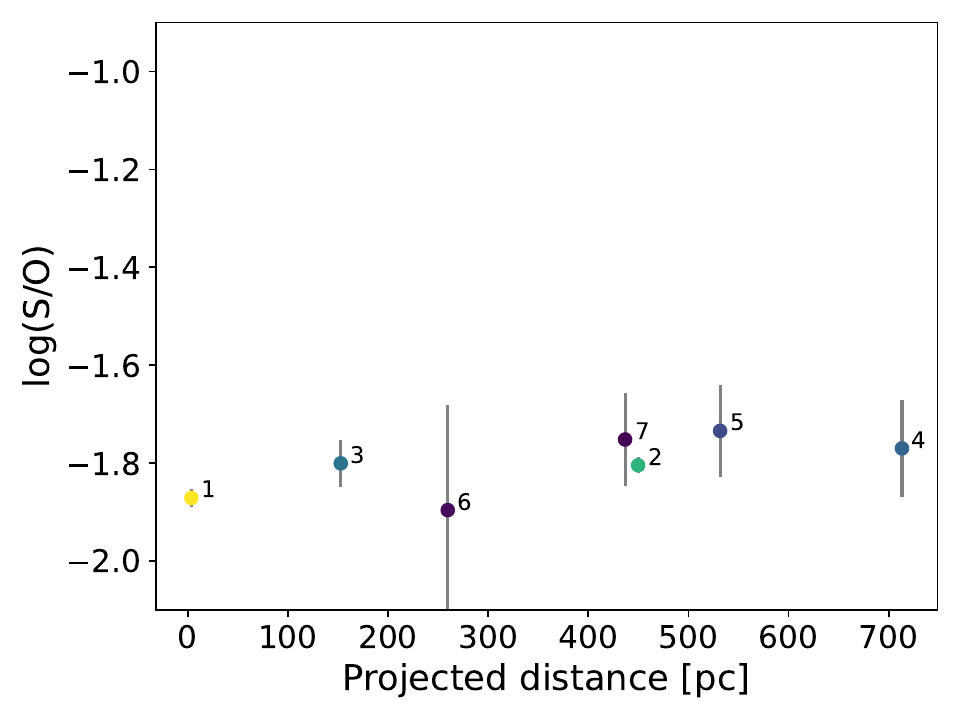}
    \includegraphics[width=0.45\textwidth]{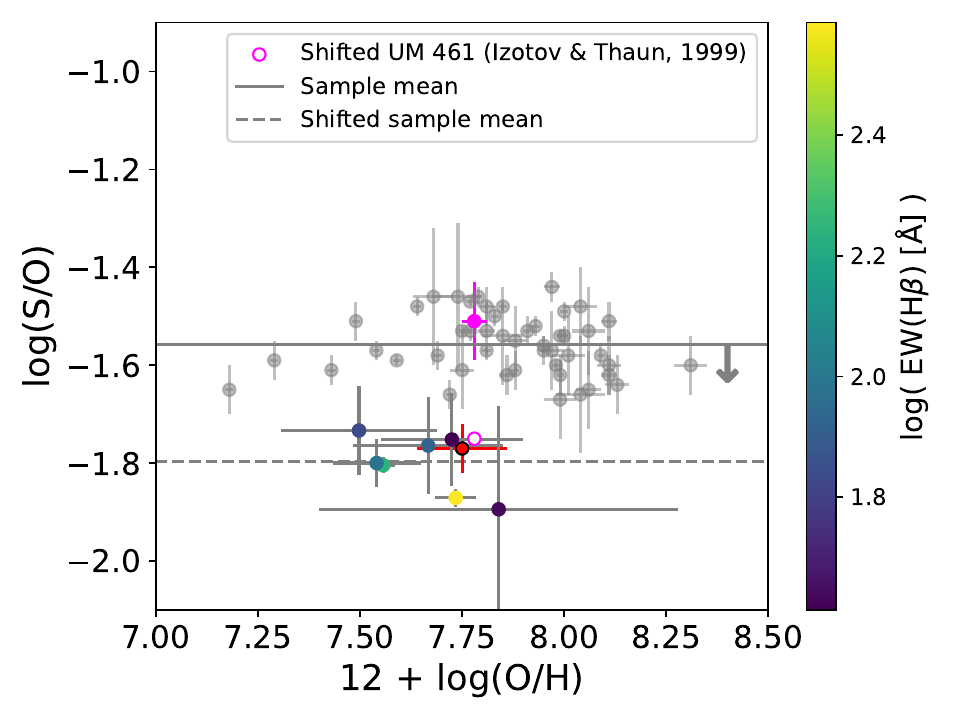}
    
      \includegraphics[width=0.45\textwidth]{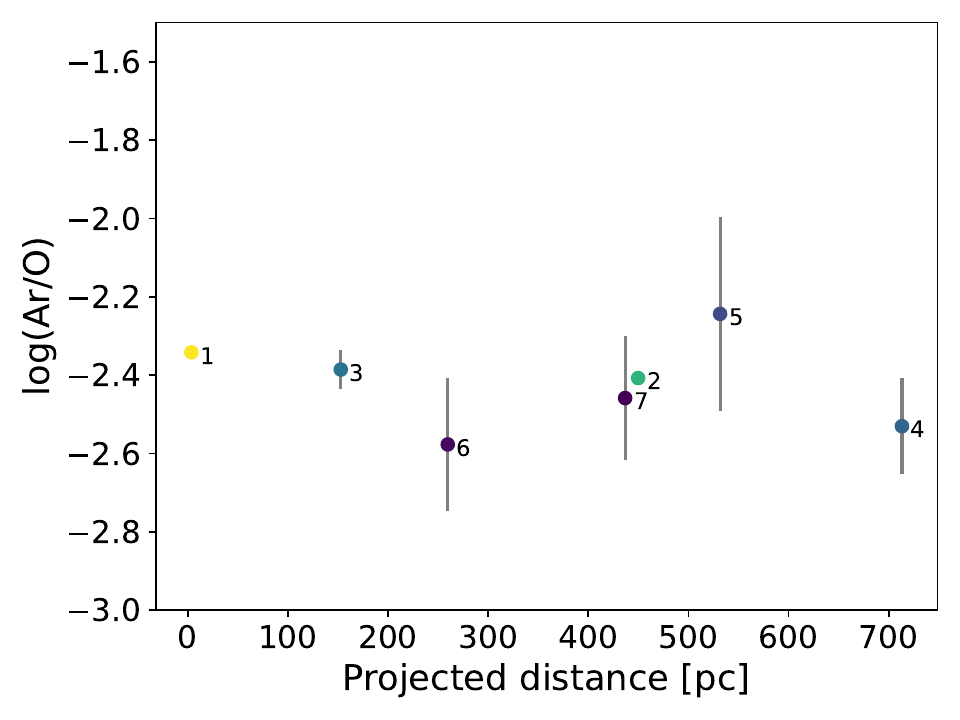}
    \includegraphics[width=0.45\textwidth]{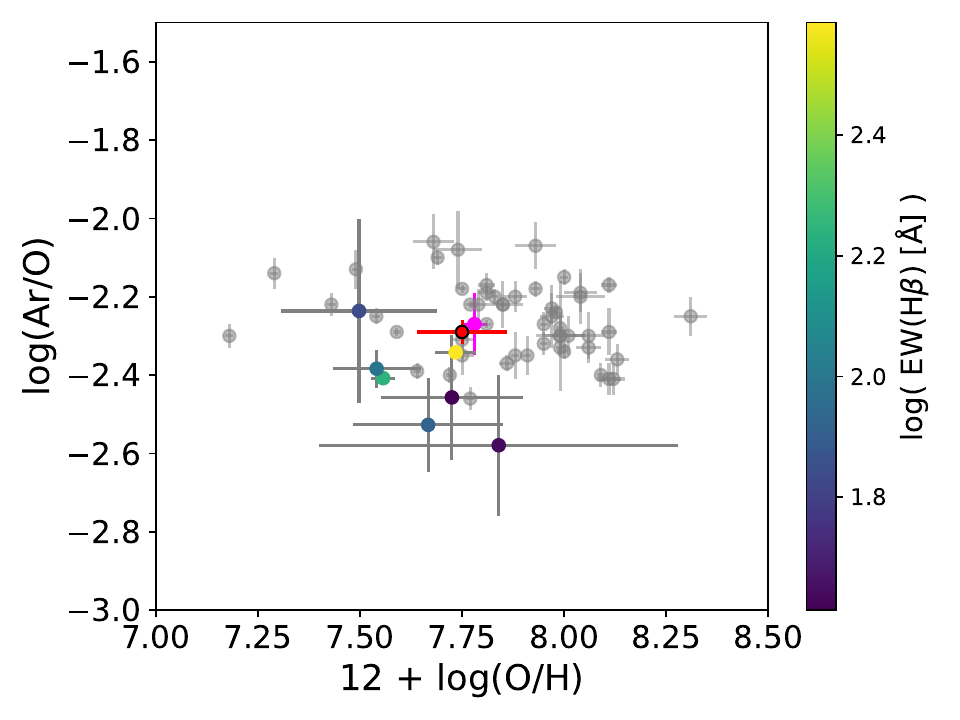}
    
    \caption{\textit{Left column:} Variation of N/O (upper panel), S/O (middle panel) and Ar/O (lower panel) with projected distance from R1 for the seven H~\textsc{ii} regions shown in Figure \ref{fig:MUSE_Ha}.  \textit{Right column:} Variation of N/O (upper panel), S/O (middle panel), and Ar/O (lower panel) with respect to the gas-phase metallicities for the seven H~\textsc{ii} regions in UM 461, color-coded with respect to EW(\hb). The gray circles denote the abundance ratios of dwarf galaxies from \citet{Izotov1999}. Their estimations for UM 461 are highlighted in magenta. In the middle panel, their S/O, recalculated using the ICF(S$^{+}+$$S^{++}$) of \citet{Dors2016}, is shown as a magenta circle. In the same panel, we indicate the original and shifted mean log(S/O) of the \citet{Izotov1999} galaxies as solid and dashed gray lines, respectively. The shifted mean represents its maximum expected change from using the ICF of \citet{Dors2016}. For reference, the solar values of these ratios from \citet{Asplund2021} are log(N/O)$_\odot=-0.86$, log(S/O)$_\odot=-1.57$, and log(Ar/O)$_\odot=-2.3$. }
    \label{fig:element_to_O_ratio}
\end{figure*}

Figure~\ref{fig:OH} (upper panel) presents the distribution of gas-phase metallicities relative to the projected distance from the brightest star-forming region (R1 or region 1), showing that there is no evidence of radial gradient in the gas-phase metallicity or any dependence on the EW(H$\beta$) (indicative of the stellar ages). We also explore the correlation between the gas-phase metallicity and the H$\alpha$ fluxes (indicative of star-formation rate) of individual H\textsc{ii} regions in Figure~\ref{fig:OH} (lower panel). F(H$\beta$) and EW(H$\beta$) are highly correlated with a Pearson correlation coefficient of $r = 0.96$ (p-value of $p=0.0004$), as seen in the lower panel of Figure~\ref{fig:OH}. This correlation does not appear to be related to their gas-phase abundances.

The left column of Figure~\ref{fig:element_to_O_ratio} displays the variation of N/O, S/O and Ar/O with respect to projected distance. Like oxygen abundance, we do not observe a radial gradient for log(N/O), log(S/O) and log(Ar/O). The log(N/O) and log(Ar/O) ratios  have Pearson correlation coefficients of $|r| < 0.3$ with respect to projected distance. The log(S/O) ratio with respect to projected distance has a coefficient of $-0.69$, but it is not statistically significant ($p=0.08$). These results suggest that the interstellar medium is chemically homogeneous. The homogeneity of log(N/O) in UM 461 was also observed by \citet{Lagos2018}. 

In the right panels of Figure~\ref{fig:element_to_O_ratio}, we compare the N/O, S/O, and Ar/O ratios versus O/H of the seven H~\textsc{ii} regions to the sample of dwarf galaxies from \citet{Izotov1999}. In general, N/O and Ar/O are consistent with the sample, but S/O is systematically lower in all H~\textsc{ii} regions. As described in \citet{Izotov1998}, the sample uses the sulfur ICF from \citet{Izotov1994}, derived from the photoionization models of \citet{Stasinska1990}. In fact, UM 461 has the highest sulfur ICF of all galaxies in their sample (2.66), as its O$^+/$O is also the lowest (0.08). Sulfur ICFs are generally accurate for O$^+/$O $>0.2$ \citep{Vermeij2002, Dors2016}, so to assess the impact of using different ICFs, we recalculate the total S abundance of \citet{Izotov1999} of UM 461 using the ICF(S$^+$ + S$^{2+}$) from \citet{Dors2016} as described in Section~\ref{sec:physical_conditions}. From here, we obtain a shift of $-0.24$ dex relative to the original value. As the original sulfur ICF is the highest in the sample, this shift is also the largest. This agrees with \citet{Dors2016} as they found the \citet{Izotov1994}\footnote{\citet{Dors2016} cite \citet{Thuan1995} as the source of the ICF. The description of the ICF is the same in both papers.} sulfur ICF yields total abundances $\sim 0.1$ dex higher than the other five ICFs from the literature for galaxies with intermediate ($7.6 < 12 + \log($O/H$) < 8.2$) metallicity.  In the middle-right panel of Figure~\ref{fig:element_to_O_ratio}, the solid gray line represents the mean of the sample of galaxies, meanwhile the dashed line is the mean shifted by $-0.24$ dex, showing the maximum possible displacement when we account for the ICF variations. This means the data points would fall between the two lines if the \citet{Dors2016} ICF were used.

This is similar to a comparison made by \citet{Berg2021}. They calculate ICFs from new photoionization models and apply them to a pair of BCDs, J104457 and J141851. They find their ICF(S$^+$ + S$^{2+}$) values are systematically lower than those obtained with the \citet{Izotov1994} relation. Using the \citet{Izotov1994} ICFs translates into an increment of log(S/O) of 0.48 dex and 0.24 dex for J141851 and J104457, respectively. \citet{Berg2021} notes differences in the atomic data of S$^{2+}$ employed by the foundational photoionization models may be the reason behind the shifts. Finally, calculating log(S/O) from the ICF((S$^+$ + S$^{2+}$)/(O$^+$ + O$^{2+}$)) of \citet{Amayo2021MNRAS.505.2361A} agrees with all the values presented in this work. Our Ar/O and Cl/O estimates mostly agree with the values obtained from the ICFs of \citet{Amayo2021MNRAS.505.2361A}, except in region 2 where log(C/O) decreases by 0.13 dex. This similar to the findings of \citet{Orte-Garcia2025} for a few highly ionized objects.

Under these considerations, our H~\textsc{ii} regions are in agreement with the distribution of the H~\textsc{ii} region sample from \citet{Izotov1994}. The shifted UM 461 value agrees with the COS-matched aperture R1. However, the individual region 1, contained in R1, has a lower log(S/O) value. The difference comes from the selection of $T_e(\mathrm{intermediate})$. Considering $T_e($\ariii) for region 1 instead of $T_e($\siii) reconciles the differences with R1.

The N/O, S/O, and Ar/O ratios appear to have anti-correlations with gas-phase metallicity ($r=-0.71, -0.67$ and $-0.69$, respectively), but have p-values of $p>0.07$ and are dependent on region 6, which has high uncertainties. Omitting this region, the anti-correlation disappears, increasing the Pearson coefficients by $0.20$ and the p-values to $p>0.30$. Our distribution of Ar/O is consistent with the solar value of \citet[][log(Ar/O$)=-2.3$]{Asplund2021}, pointing to a constant Ar/O versus O/H relation, similar to other works in the literature \citep[e.g.,][]{Izotov2006,MirandaPerez2023ApJ...952...76M}. The sub-solar log(S/O) of our regions are consistent with other star-forming galaxies of similar metallicity \citep[e.g.,][]{MirandaPerez2023ApJ...952...76M,Scholte2026MNRAS.551g1381S}, however, this may be due on the selection of the ICF as discussed earlier in the section.

We should note that using the higher $T_e$([N~\textsc{ii}]) from the $T_e$([N~\textsc{ii}]) versus $T_e$(\siii) relation from \citet{OrteGarcia2026A&A...711A.165O} results in the increment of the median values of most abundance ratios by $\leq0.25$ dex. These ratios are still in agreement within $2\sigma$, with the exception of log(N/O) and log(S/O) in regions 1 and 2. However, considering that the resulting increments are similar for all regions, our result of the absence of a projected gradient in the metallicity and abundance ratios still holds and indicates homogeneity.

\subsection{Escape of ionizing photons}

\citet{Schaerer2022A&A...658L..11S} showed that strong Lyman Continuum (LyC) leakers (f$_{\rm{esc}}>0.1$) present a \civ\ $\lambda\lambda 1548,50/$\ciiisemi\ $\lambda\lambda 1908$ (C43) line ratio of C43 $>0.75$. They observed a high C43 ratio in the three strong LyC leakers of their sample and further justified from this criterion based on density-bounded photoionization models. Those models present a correlation between LyC escape fraction and the C43 ratio. In UM 461 R1, we find C43$=0.25\pm 0.01$, suggesting a low escape fraction of ionizing photons.

LyC leakers have also been associated with low values of the observed UV continuum slope at 1550 \AA\ \citep[$\beta_{\rm{obs}}^{1550}$,][]{Chisholm2022}. We calculate $\beta_{\rm{obs}}^{1550}= -0.78\pm0.16 $ for UM 461 R1, considering the spectral windows of \citet{Calzetti1994} in the wavelength range of $\sim$1410-1580 \AA. Using Eq. 11 of \citet{Chisholm2022}, we find $f_{\rm{esc,LyC}}\sim10^{-3}$, also suggesting a negligible escape of LyC photons. 

However, this contrasts with the high \oiii\ 88 $\mu$m/\cii\ 157 $\mu$m ratio of UM 461 which hints at considerable escape of ionizing photons. \citet{Ramambason2022} find a escape fraction of $\sim 45\%$ by modeling the IR fluxes of \citet{Cormier2015}. While $\beta_{\rm{obs}}^{1550}$ anticorrelates with ionizing photon escape, this trend has a significant scatter \citep[see Figure 8 of][]{Choustikov2024MNRAS.529.3751C}. Thus, we can not rule out a high escape fraction for individual objects, like UM 461 and Pox 186, with this indicator. \citet{Schaerer2022A&A...658L..11S} also report one object with high C43 but low escape fraction, suggesting this indicator is not sufficient in isolation to establish LyC escape. Additionally, the escape can also occur in ionization-bounded regions with holes. \citet{Katz2023MNRAS.518..270K} shows that a \cii deficit denotes the presence of LyC leaking. At the same time, they suggest objects with escape fractions $>20$\% undergoing extreme bursts of star-formation are similar to ionization-bounded regions with holes rather than density-bounded regions. The similar results obtained for UM 461 and Pox 186 \citep{Kumari2024} hint that the C43 line ratio and $\beta_{\rm{obs}}^{1550}$ are not sufficient to classify an object as a LyC leaker.

\subsection{Comparison of UM 461 and Pox 186}

The local dwarf galaxies UM 461 and Pox 186, selected for their high \oiii\ 88 $\mu$m/\cii 157 $\mu$m line ratio, possess multiple similarities including low gas-phase metallically ($12+\log($O/H$)\sim7.8$), highly ionized gas with detections of the nebular UV \oiiiuv, \civ, and \ciiisemi\ doublets, as well as the presence of \heii\ recombination lines. Evidence of outflows is also absent for both UM 461 (see Section~\ref{sec:uv_line_profiles}) and Pox 186 \citep{Kumari2024}. They also exhibit weak \civ\ P-Cygni profiles, absence of WR signatures, a visible Paschen jump arising from nebular continuum\footnote{We inspected the unpublished archival MUSE data (Program ID: 1104.A-0026(E)) of Pox 186 to find the Paschen jump.}, and high EW(\ciiisemi) ($\gtrsim 18$ \AA). High EW(\ciiisemi) and FIR \oiii\ /\cii ratios are characteristic of EoR galaxies \citep{Kumari2024,Kumari2026arXiv260930003K} and establish UM 461 and Pox 186 as their local analogues. Like Pox 186, UM 461 also follows the general trend in the EW(\ciiisemi) versus EW(\oiii\ $\lambda\lambda 4959,~5007 +$H$\beta)$ of star-forming galaxies in the local Universe \citep[see Figure 6 of][]{Kumari2024}, as the COS-matched aperture R1 exhibits an EW(\oiii+H$\beta) \sim 2870$ \AA.

The nebular emission of UM 461 R1 and Pox 186 arises from low density ($n_e\lesssim200$ cm$^{-3}$) and high temperature ($T_e($\oiii$)\sim 16000$ K) regions ionized by star formation. This also results in the optical and UV line ratios being alike, situating both galaxies in comparable regions on the emission line diagnostic diagrams (see Section~\ref{sec:diagnostic_diagrams}).

There exist certain differences between both galaxies. The equivalent widths of nebular UV detected by \citet{Kumari2024} in Pox 186 are systematically higher, with EW(\civ\ $\lambda 1548$), EW(\civ\ $\lambda 1550$) and EW(\ciiisemi\ $\lambda 1908$) $\sim 2$ times those of UM 461 R1. In UM 461 R1, we also do not detect the double-peaked \civ\ profile seen in Pox 186.

The C43 line ratios and $\beta_{\rm{obs}}^{1550}$ suggest both galaxies should not be LyC leakers, however, this contrasts with the high escape fraction expected from the high \oiii\ /\cii ratios and the models of \citet{Ramambason2022}. A larger sample of galaxies with similar characteristics to UM 461 and Pox 186 is required to further investigate this issue.

\section{Conclusions \label{sec:conclusions}}

We analyze the properties of the highly ionized local dwarf galaxy UM 461 by using new \textit{HST}/COS FUV and NUV observations in conjunction with ancillary optical, mid-infrared and far-infrared data from VLT/MUSE, \textit{Spitzer} and \textit{Herschel}, respectively. Our key results are summarized as follows:

\begin{enumerate}[i.]
    \item The COS data of the brightest star-forming region (R1) of UM 461 presents a high EW(\ciiisemi) ($18.46\pm1.03$ \AA). This is consistent with the relatively high FIR \oiii/\cii ratio ($3.6\pm0.1$) observed by \citet{Cormier2015}.
    \item At low metallicity ($Z\lesssim 0.008$), UV diagnostic diagrams are more useful than the traditional optical diagrams to distinguish galaxies dominated by star-formation to those with an AGN. We recommend using UV diagnostics involving ratios with \siiii\ and \ciiisemi\ lines like \ciiisemi/\heii, \siiii/\heii, and \siiii/\ciiisemi, when observed, for distinguishing the ionization source.
    \item The aperture-matched VLT/MUSE data shows nebular \heii\ $\lambda 4686$ emission in UM 461. Photoionization models with a standard IMF lacking stars with masses beyond $\sim 100~M_\odot$ cannot reproduce the \heii/H$\beta$ and \nii/H$\alpha$ ratios simultaneously, suggesting the presence of Very Massive Stars. However, aperture effects on the \heii\ line \citep{GonzalezTora2025} may also suggest weak-winded WNE stars may be present but undetected instead.
    \item We identify seven H~\textsc{ii} regions using VLT/MUSE optical data with \texttt{ifuanal}. We do not observe a projected gradient in either their metallicities or in their N/O, S/O, and Ar/O ratios.
    \item We propose a new curve in the BPT diagram (Equation~\ref{eq:civ_max_line}) that indicates an upper limit to the star-forming galaxies in which nebular \civ~ should be detected.
    \item UV tracers of LyC leakage indicate $<10\%$ of ionizing photons escape from UM 461. However, a higher escape fraction is expected from its high \oiii/\cii FIR line ratio and modeling of IR lines \citep{Ramambason2022} from the literature. This contrast is also observed in the pilot study of Pox 186 \citep{Kumari2024}, supporting the questions raised on the usage of the observed UV continuum slope $\beta_{\rm{obs}}^{1550}$ and the \civ/\ciiisemi~line ratio to identify LyC leakers.
\end{enumerate}

%% Please use the acknowledgment and contribution environments. This will 
%% be anonomyized when the "anonymous" style option is used. 
\begin{acknowledgments}
This research is based on observations made with the NASA/ESA Hubble Space Telescope obtained from the Space Telescope Science Institute, which is operated by the Association of Universities for Research in Astronomy, Inc., under NASA contract NAS 5–26555. These observations are associated with program 17729.

Some of the data presented in this article were obtained from the Mikulski Archive for Space Telescopes (MAST) at the Space Telescope Science Institute. The specific observations analyzed can be accessed via \dataset[doi:10.17909/nhzp-sy97]{https://doi.org/10.17909/nhzp-sy97}.

Based on observations collected at the European Organisation for Astronomical Research in the Southern Hemisphere under ESO programme 0104.D-0199, and/or processed data created thereof.

The authors thank the anonymous referee for their constructive comments. OA acknowledges support from CONAHCyT Beca Nacional para estudios de Posgrado, CVU 1142736. OA and AW acknowledge support of grant UNAM/DGAPA/PAPIIT IN107725. JW gratefully acknowledges support from the Cosmic Dawn Center through the DAWN Fellowship. The Cosmic Dawn Center (DAWN) is funded by the Danish National Research Foundation under grant No. 140. CM acknowledges the support of grants UNAM/DGAPA/PAPIIT IN113226.
\end{acknowledgments}

\facilities{HST(COS and WFC3), VLT:Yepun(MUSE)}

\software{astropy \citep{Astropy2013,Astropy2018,Astropy2022}, 
          CALCOS (STScI),
          dust\_extinction \citep{Gordon2024},
          HII-CHI-mistry-IR \citep{PerezDiaz2024},
          ifuanal \citep{Lyman2018},
          jupyter \citep{Jupyter2016},
          PyNeb \citep{Luridiana2015},
          SpectRes \citep{Carnall2017arXiv},
          STARLIGHT \citep{CidFernandes2005,CidFernades2011}.
          }

\appendix

\section{Atomic data references \label{sec:atomic_data}}

The references of the transitional probabilities and collision strengths considered in this work are listed on Table~\ref{tab:atomic_data}.

\begin{deluxetable}{lcc}[h]
%\digitalasset
\tablewidth{0pt}
\tablecaption{Atomic data references \label{tab:atomic_data}}
\tablehead{
\colhead{Ion} & \colhead{Transition probabilities} & \colhead{Collision strengths}
}
\startdata
H$^{0}$ & \citet{Storey1995} & - \\
C$^{2+}$ & \citet{Wiese1996} & \citet{Berrington1985} \\
C$^{3+}$ & \citet{Wiese1996} & \citet{Aggarwal2004} \\
N$^{+}$ & \citet{FroeseFischer2004} & \citet{Tayal2011} \\
O$^{+}$ & \citet{Zeippen1982} & \citet{Kisielus2009} \\
O$^{2+}$ & \citet{Storey2000} & \citet{Tayal2017} \\
Si$^{2+}$ & \citet{Mendoza1983,Ojha1988} & \citet{Dufton1989} \\
S$^{+}$ & \citet{Rynkun2019} & \citet{Tayal2010} \\
S$^{2+}$ & \citet{FroeseFischer2006} & \citet{Tayal1999}  \\
Cl$^{2+}$ & \citet{Rynkun2019} & \citet{Butler1989} \\
Cl$^{3+}$ & \citet{Kaufman1986} & \citet{Galavis1995} \\
Ar$^{2+}$ & \citet{MunozBurgos2009} & \citet{MunozBurgos2009} \\
Ar$^{3+}$ & \citet{Rynkun2019} & \citet{Ramsbottom1997} \\
\enddata
\end{deluxetable}

\section{Electron Density from Si III] doublet}
\begin{figure}
    \centering
    \includegraphics[width=0.5\linewidth]{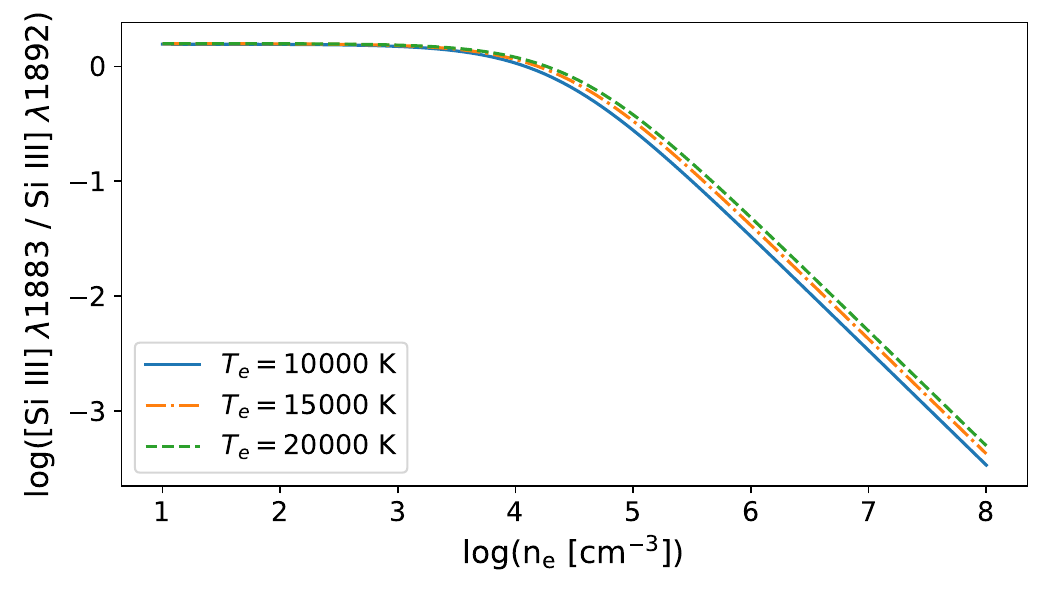}
    \caption{Variation of [\siiii\ $\lambda$1883/\siiii\ $\lambda$1892 emissivity ratio with respect to electron density, $n_e$, and temperature, $T_e$.}
    \label{fig:Si3_doublet_ratio}
\end{figure}

Figure~\ref{fig:Si3_doublet_ratio} shows the variation of \siiii\ $\lambda$1883/1893 emissivity ratio with respect to electron density at different electron temperatures. 
\label{sec:Ne_SiIII}

\section{Individual H~\textsc{ii} region properties}

\begin{figure*}
    \centering
    \includegraphics[width=\linewidth]{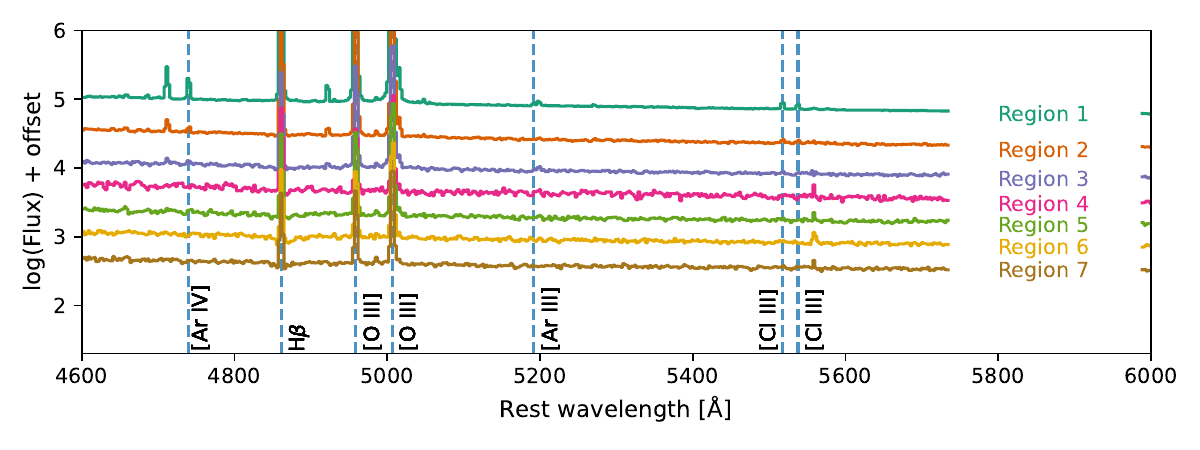}
    \includegraphics[width=\linewidth]{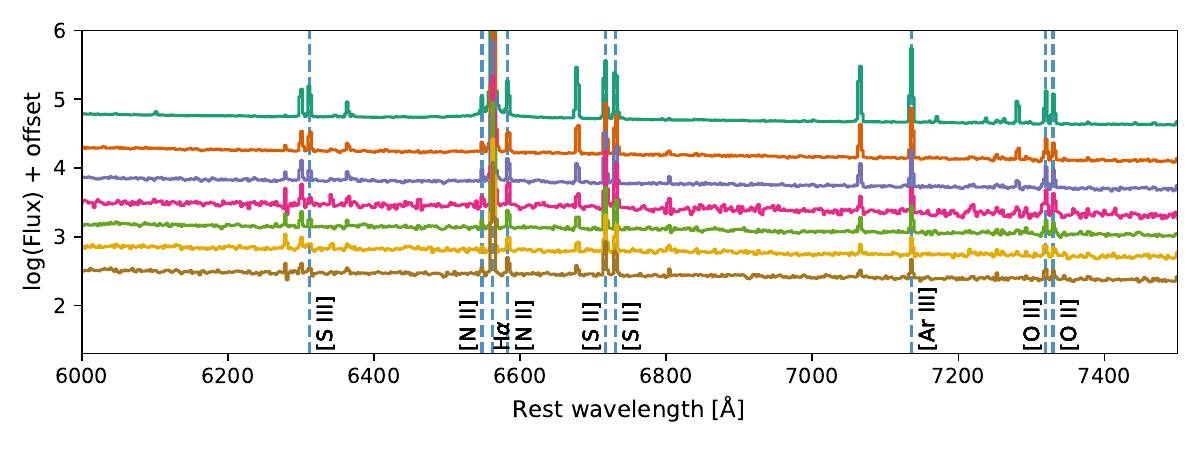}
    \includegraphics[width=\linewidth]{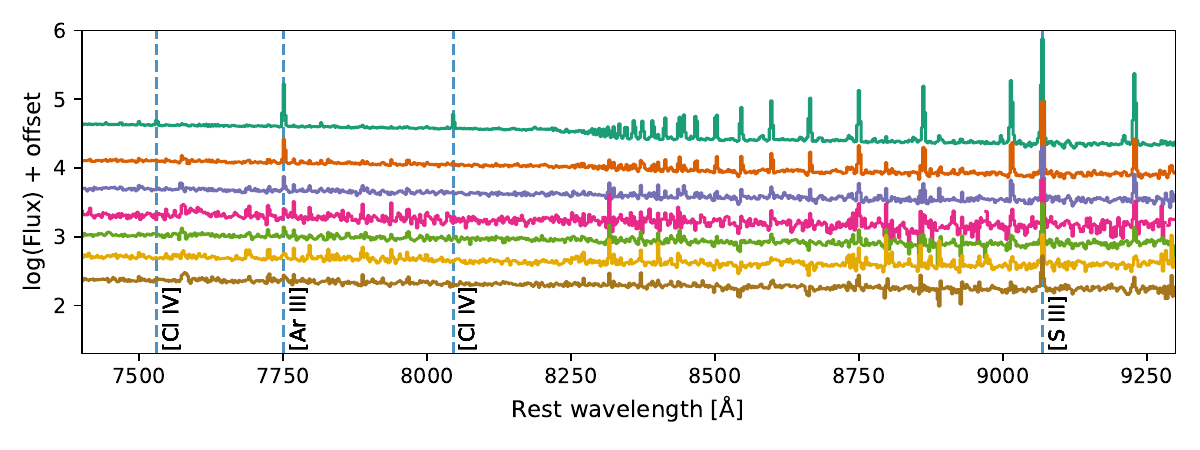}
    \caption{Optical spectra of the individual regions in the wavelength range of 4600-6000 \AA\ (top panel), 6000-7500 \AA\ (middle panel), and  7200-9300 \AA\ (bottom panel). Observed lines used for calculating intrinsic extinction and ionic abundances are marked with dashed lines. Spectra are rebbined by 2 pixels for better visibility.}
    \label{fig:ifuanal_spectra}
\end{figure*}

Tables~\ref{tab:individual_hii_region_fluxes} and $\ref{tab:individual_hii_region_fluxes_2}$ show the observed and intrinsic emission line fluxes of the H~\textsc{ii} regions identified with \texttt{ifuanal}. In Table~\ref{tab:individual_hii_region_abundances}, we list the estimated ionic temperatures, ionic abundances, ICFs, and total abundances of the individual H~\textsc{ii} regions. Spectra of the individual regions are shown in Figure~\ref{fig:ifuanal_spectra}.

\begin{sidewaystable}
\centering
\caption{Observed (F$_\lambda$) and intrinsic (I$_\lambda$) emission line fluxes of H~\textsc{ii} regions 1-4 (relative to H$\beta$=100). F(H$\beta$) in units of $10^{-15}$ erg s$^{-1}$ cm$^{-2}$. \label{tab:individual_hii_region_fluxes}}
\begin{tabular}{lcccccccccc}

\toprule
Line & Wavelength [\AA] & F$_{\lambda,1}$ & I$_{\lambda,1}$ & F$_{\lambda,2}$ & I$_{\lambda,2}$ & F$_{\lambda,3}$ & I$_{\lambda,3}$ & F$_{\lambda,4}$ & I$_{\lambda,4}$ \\

\midrule
$[$Fe~\textsc{iii}$]$ & 4658.10 & 0.15 $\pm$ 0.02 & 0.15 $\pm$ 0.02 & 0.36 $\pm$ 0.06 & 0.36 $\pm$ 0.07 & - & - & - & - \\
He~\textsc{ii} & 4685.68 & 0.14 $\pm$ 0.02 & 0.14 $\pm$ 0.02 & - & - & - & - & - & - \\
$[$Ar~\textsc{iv}$]$+He~\textsc{i} & 4711.33 & 2.44 $\pm$ 0.12 & 2.45 $\pm$ 0.12 & 1.57 $\pm$ 0.03 & 1.58 $\pm$ 0.03 & 1.20 $\pm$ 0.25 & 1.20 $\pm$ 0.25 & - & - \\
$[$Ar~\textsc{iv}$]$ & 4740.20 & 1.47 $\pm$ 0.04 & 1.48 $\pm$ 0.04 & 0.60 $\pm$ 0.08 & 0.61 $\pm$ 0.08 & 0.81 $\pm$ 0.24 & 0.81 $\pm$ 0.24 & - & - \\
H$\beta$ & 4861.36 & 100.00 $\pm$ 2.83 & 100.00 $\pm$ 2.83 & 100.00 $\pm$ 1.19 & 100.00 $\pm$ 1.19 & 100.00 $\pm$ 1.38 & 100.00 $\pm$ 1.38 & 100.00 $\pm$ 1.44 & 100.00 $\pm$ 1.44 \\
He~\textsc{i} & 4921.93 & 0.73 $\pm$ 0.16 & 0.73 $\pm$ 0.16 & 0.66 $\pm$ 0.14 & 0.66 $\pm$ 0.14 & 0.47 $\pm$ 0.26 & 0.47 $\pm$ 0.26 & - & - \\
$[$O~\textsc{iii}$]$ & 4958.91 & 218.30 $\pm$ 4.86 & 217.52 $\pm$ 4.84 & 184.28 $\pm$ 3.62 & 183.57 $\pm$ 3.61 & 144.49 $\pm$ 2.42 & 143.97 $\pm$ 2.41 & 60.23 $\pm$ 1.05 & 59.74 $\pm$ 1.04 \\
$[$O~\textsc{iii}$]$ & 5006.84 & 598.36 $\pm$ 31.85 & 595.19 $\pm$ 31.68 & 547.46 $\pm$ 10.90 & 544.37 $\pm$ 10.83 & 427.58 $\pm$ 6.47 & 425.28 $\pm$ 6.44 & 180.40 $\pm$ 3.71 & 178.24 $\pm$ 3.67 \\
$[$Ar~\textsc{iii}$]$ & 5191.82 & 0.09 $\pm$ 0.01 & 0.09 $\pm$ 0.01 & - & - & - & - & - & - \\
$[$N~\textsc{i}$]$ & 5199.00 & 0.19 $\pm$ 0.03 & 0.19 $\pm$ 0.03 & 0.13 $\pm$ 0.04 & 0.12 $\pm$ 0.04 & 0.68 $\pm$ 0.09 & 0.67 $\pm$ 0.09 & 0.66 $\pm$ 0.53 & 0.64 $\pm$ 0.52 \\
$[$Cl~\textsc{iii}$]$ & 5517.71 & 0.25 $\pm$ 0.01 & 0.24 $\pm$ 0.01 & 0.26 $\pm$ 0.02 & 0.25 $\pm$ 0.02 & 0.21 $\pm$ 0.13 & 0.21 $\pm$ 0.12 & - & - \\
$[$Cl~\textsc{iii}$]$ & 5537.87 & 0.17 $\pm$ 0.01 & 0.17 $\pm$ 0.01 & 0.21 $\pm$ 0.04 & 0.21 $\pm$ 0.04 & - & - & - & - \\
$[$O~\textsc{i}$]$ & 6300.30 & 1.21 $\pm$ 0.03 & 1.16 $\pm$ 0.03 & 1.36 $\pm$ 0.04 & 1.31 $\pm$ 0.04 & 2.71 $\pm$ 0.08 & 2.61 $\pm$ 0.08 & 3.39 $\pm$ 0.14 & 3.11 $\pm$ 0.13 \\
$[$S~\textsc{iii}$]$ & 6312.06 & 1.43 $\pm$ 0.04 & 1.38 $\pm$ 0.04 & 1.60 $\pm$ 0.05 & 1.54 $\pm$ 0.05 & 1.44 $\pm$ 0.16 & 1.39 $\pm$ 0.16 & 1.12 $\pm$ 0.23 & 1.02 $\pm$ 0.21 \\
$[$O~\textsc{i}$]$ & 6363.78 & 0.40 $\pm$ 0.01 & 0.39 $\pm$ 0.01 & 0.48 $\pm$ 0.04 & 0.46 $\pm$ 0.04 & 0.95 $\pm$ 0.06 & 0.91 $\pm$ 0.06 & 1.35 $\pm$ 0.17 & 1.23 $\pm$ 0.15 \\
$[$N~\textsc{ii}$]$ & 6548.05 & 0.58 $\pm$ 0.04 & 0.55 $\pm$ 0.04 & 0.59 $\pm$ 0.02 & 0.56 $\pm$ 0.02 & 1.24 $\pm$ 0.11 & 1.19 $\pm$ 0.10 & 1.30 $\pm$ 0.39 & 1.18 $\pm$ 0.35 \\
H$\alpha$ & 6562.85 & 298.52 $\pm$ 12.64 & 286.07 $\pm$ 12.11 & 299.35 $\pm$ 2.81 & 286.08 $\pm$ 2.68 & 298.73 $\pm$ 4.27 & 286.07 $\pm$ 4.09 & 315.27 $\pm$ 4.63 & 286.17 $\pm$ 4.21 \\
$[$N~\textsc{ii}$]$ & 6583.45 & 1.62 $\pm$ 0.04 & 1.55 $\pm$ 0.04 & 1.89 $\pm$ 0.05 & 1.81 $\pm$ 0.05 & 3.85 $\pm$ 0.09 & 3.69 $\pm$ 0.08 & 5.67 $\pm$ 0.21 & 5.14 $\pm$ 0.19 \\
He~\textsc{i} & 6678.15 & 3.28 $\pm$ 0.09 & 3.13 $\pm$ 0.08 & 3.04 $\pm$ 0.08 & 2.89 $\pm$ 0.08 & 2.85 $\pm$ 0.07 & 2.72 $\pm$ 0.06 & 2.38 $\pm$ 0.14 & 2.15 $\pm$ 0.13 \\
$[$S~\textsc{ii}$]$ & 6716.44 & 4.40 $\pm$ 0.10 & 4.20 $\pm$ 0.09 & 5.96 $\pm$ 0.11 & 5.68 $\pm$ 0.10 & 11.78 $\pm$ 0.13 & 11.25 $\pm$ 0.12 & 18.00 $\pm$ 0.41 & 16.24 $\pm$ 0.37 \\
$[$S~\textsc{ii}$]$ & 6730.82 & 3.42 $\pm$ 0.10 & 3.27 $\pm$ 0.10 & 4.30 $\pm$ 0.08 & 4.10 $\pm$ 0.08 & 8.33 $\pm$ 0.17 & 7.95 $\pm$ 0.16 & 12.42 $\pm$ 0.60 & 11.20 $\pm$ 0.54 \\
He~\textsc{i} & 7065.19 & 3.78 $\pm$ 0.10 & 3.59 $\pm$ 0.09 & 2.34 $\pm$ 0.04 & 2.21 $\pm$ 0.04 & 2.58 $\pm$ 0.05 & 2.45 $\pm$ 0.05 & 1.99 $\pm$ 0.26 & 1.77 $\pm$ 0.23 \\
$[$Ar~\textsc{iii}$]$ & 7135.80 & 5.66 $\pm$ 0.15 & 5.37 $\pm$ 0.14 & 5.45 $\pm$ 0.06 & 5.16 $\pm$ 0.06 & 5.05 $\pm$ 0.09 & 4.78 $\pm$ 0.09 & 4.28 $\pm$ 0.16 & 3.80 $\pm$ 0.14 \\
He~\textsc{i} & 7281.35 & 0.73 $\pm$ 0.02 & 0.69 $\pm$ 0.02 & 0.60 $\pm$ 0.03 & 0.57 $\pm$ 0.02 & 0.69 $\pm$ 0.13 & 0.66 $\pm$ 0.12 & 0.66 $\pm$ 0.32 & 0.58 $\pm$ 0.28 \\
$[$O~\textsc{ii}$]$ & 7318.92 & 1.04 $\pm$ 0.03 & 0.98 $\pm$ 0.03 & 1.18 $\pm$ 0.14 & 1.12 $\pm$ 0.13 & 2.19 $\pm$ 0.30 & 2.07 $\pm$ 0.28 & 3.35 $\pm$ 1.20 & 2.96 $\pm$ 1.06 \\
$[$O~\textsc{ii}$]$ & 7329.66 & 0.87 $\pm$ 0.02 & 0.82 $\pm$ 0.02 & 0.91 $\pm$ 0.01 & 0.86 $\pm$ 0.01 & 1.82 $\pm$ 0.03 & 1.72 $\pm$ 0.03 & 3.04 $\pm$ 0.09 & 2.68 $\pm$ 0.08 \\
$[$Cl~\textsc{iv}$]$ & 7530.54 & 0.10 $\pm$ 0.01 & 0.10 $\pm$ 0.01 & - & - & - & - & - & - \\
$[$Ar~\textsc{iii}$]$ & 7751.06 & 1.40 $\pm$ 0.04 & 1.32 $\pm$ 0.04 & 1.30 $\pm$ 0.04 & 1.21 $\pm$ 0.04 & 1.06 $\pm$ 0.10 & 1.00 $\pm$ 0.09 & 0.85 $\pm$ 0.41 & 0.74 $\pm$ 0.36 \\
$[$Cl~\textsc{iv}$]$ & 8045.62 & 0.26 $\pm$ 0.01 & 0.25 $\pm$ 0.01 & - & - & - & - & - & - \\
$[$S~\textsc{iii}$]$ & 9069.00 & 8.61 $\pm$ 0.21 & 7.97 $\pm$ 0.20 & 8.02 $\pm$ 0.13 & 7.38 $\pm$ 0.12 & 7.07 $\pm$ 0.40 & 6.54 $\pm$ 0.37 & 7.73 $\pm$ 0.35 & 6.48 $\pm$ 0.30 \\
\midrule
F(H$\beta$) & - & 66.59 $\pm$ 1.33 & 75.53 $\pm$ 1.51 & 8.24 $\pm$ 0.07 & 9.42 $\pm$ 0.08 & 2.97 $\pm$ 0.03 & 3.37 $\pm$ 0.03 & 0.76 $\pm$ 0.01 & 1.01 $\pm$ 0.01 \\
$E(B-V)$ & - & 0.07 $\pm$ 0.04 & - & 0.07 $\pm$ 0.01 & - & 0.07 $\pm$ 0.01 & - & 0.12 $\pm$ 0.01 & - \\
\bottomrule
\end{tabular}
\end{sidewaystable}

\begin{sidewaystable}
    \caption{Emission line fluxes of H~\textsc{ii} regions 5, 6 \& 7 (relative to H$\beta$=100). F(H$\beta$) in units of $10^{-15}$ erg s$^{-1}$ cm$^{-2}$. \label{tab:individual_hii_region_fluxes_2}}
    \begin{tabular}{lcccccccc}
    \toprule
    Line & Wavelength [\AA] & F$_{\lambda,5}$ & I$_{\lambda,5}$ & F$_{\lambda,6}$ & I$_{\lambda,6}$ & F$_{\lambda,7}$ & I$_{\lambda,7}$ \\

    \midrule
         $[$Fe~\textsc{iii}$]$ & 4658.10 & 1.43 $\pm$ 0.92 & 1.45 $\pm$ 0.93 & - & - & 2.70 $\pm$ 1.31 & 2.73 $\pm$ 1.32 \\
         He~\textsc{ii} & 4685.68 & - & - & - & - & - & - \\
$[$Ar~\textsc{iv}$]$+He~\textsc{i} & 4711.33 & 0.81 $\pm$ 0.29 & 0.82 $\pm$ 0.30 & 0.99 $\pm$ 0.31 & 1.00 $\pm$ 0.32 & 1.63 $\pm$ 1.26 & 1.64 $\pm$ 1.26 \\
$[$Ar~\textsc{iv}$]$ & 4740.20 & 1.66 $\pm$ 1.47 & 1.67 $\pm$ 1.48 & - & - & - & - \\
H$\beta$ & 4861.36 & 100.00 $\pm$ 3.12 & 100.00 $\pm$ 3.12 & 100.00 $\pm$ 3.79 & 100.00 $\pm$ 3.79 & 100.00 $\pm$ 3.32 & 100.00 $\pm$ 3.33 \\
$[$O~\textsc{iii}$]$ & 4958.91 & 136.63 $\pm$ 3.68 & 135.80 $\pm$ 3.66 & 117.46 $\pm$ 3.79 & 116.56 $\pm$ 3.76 & 126.22 $\pm$ 3.05 & 125.65 $\pm$ 3.04 \\
$[$O~\textsc{iii}$]$ & 5006.84 & 409.97 $\pm$ 10.63 & 406.30 $\pm$ 10.53 & 345.48 $\pm$ 9.60 & 341.59 $\pm$ 9.50 & 383.74 $\pm$ 11.60 & 381.19 $\pm$ 11.52 \\
$[$Ar~\textsc{iii}$]$ & 5191.82 & - & - & - & - & - & - \\
$[$N~\textsc{i}$]$ & 5199.00 & 0.57 $\pm$ 0.20 & 0.55 $\pm$ 0.20 & - & - & - & - \\
$[$Cl~\textsc{iii}$]$ & 5517.71 & 0.32 $\pm$ 0.25 & 0.31 $\pm$ 0.25 & - & - & - & - \\
$[$Cl~\textsc{iii}$]$ & 5537.87 & - & - & - & - & - & - \\
$[$O~\textsc{i}$]$ & 6300.30 & 2.89 $\pm$ 0.30 & 2.71 $\pm$ 0.29 & 3.09 $\pm$ 0.50 & 2.85 $\pm$ 0.47 & 3.00 $\pm$ 0.75 & 2.86 $\pm$ 0.72 \\
$[$S~\textsc{iii}$]$ & 6312.06 & 1.63 $\pm$ 0.36 & 1.53 $\pm$ 0.34 & 1.35 $\pm$ 0.77 & 1.25 $\pm$ 0.71 & 1.58 $\pm$ 0.30 & 1.51 $\pm$ 0.29 \\
$[$O~\textsc{i}$]$ & 6363.78 & 0.89 $\pm$ 0.21 & 0.83 $\pm$ 0.20 & 1.18 $\pm$ 0.34 & 1.09 $\pm$ 0.31 & 1.29 $\pm$ 0.29 & 1.23 $\pm$ 0.27 \\
$[$N~\textsc{ii}$]$ & 6548.05 & 1.16 $\pm$ 0.33 & 1.08 $\pm$ 0.30 & 0.95 $\pm$ 0.69 & 0.87 $\pm$ 0.63 & 1.79 $\pm$ 0.48 & 1.70 $\pm$ 0.46 \\
H$\alpha$ & 6562.85 & 307.48 $\pm$ 7.91 & 286.12 $\pm$ 7.36 & 313.40 $\pm$ 8.51 & 286.16 $\pm$ 7.77 & 301.81 $\pm$ 8.45 & 286.09 $\pm$ 8.01 \\
$[$N~\textsc{ii}$]$ & 6583.45 & 4.29 $\pm$ 0.21 & 3.99 $\pm$ 0.20 & 5.14 $\pm$ 0.43 & 4.69 $\pm$ 0.39 & 6.25 $\pm$ 0.26 & 5.92 $\pm$ 0.25 \\
He~\textsc{i} & 6678.15 & 2.77 $\pm$ 0.19 & 2.57 $\pm$ 0.17 & 2.95 $\pm$ 0.09 & 2.68 $\pm$ 0.08 & 2.92 $\pm$ 0.35 & 2.76 $\pm$ 0.33 \\
$[$S~\textsc{ii}$]$ & 6716.44 & 13.11 $\pm$ 0.46 & 12.14 $\pm$ 0.42 & 15.47 $\pm$ 0.69 & 14.05 $\pm$ 0.63 & 16.94 $\pm$ 0.49 & 16.01 $\pm$ 0.46 \\
$[$S~\textsc{ii}$]$ & 6730.82 & 9.16 $\pm$ 0.30 & 8.48 $\pm$ 0.28 & 10.70 $\pm$ 0.50 & 9.71 $\pm$ 0.45 & 12.16 $\pm$ 0.42 & 11.49 $\pm$ 0.39 \\
He~\textsc{i} & 7065.19 & 2.45 $\pm$ 0.23 & 2.25 $\pm$ 0.21 & 2.21 $\pm$ 0.22 & 1.98 $\pm$ 0.20 & 2.05 $\pm$ 0.32 & 1.92 $\pm$ 0.30 \\
$[$Ar~\textsc{iii}$]$ & 7135.80 & 5.44 $\pm$ 0.16 & 4.98 $\pm$ 0.15 & 4.59 $\pm$ 0.42 & 4.11 $\pm$ 0.37 & 5.38 $\pm$ 0.28 & 5.04 $\pm$ 0.26 \\
He~\textsc{i} & 7281.35 & 0.58 $\pm$ 0.32 & 0.53 $\pm$ 0.29 & 0.70 $\pm$ 0.50 & 0.62 $\pm$ 0.44 & - & - \\
$[$O~\textsc{ii}$]$ & 7318.92 & 1.99 $\pm$ 0.58 & 1.81 $\pm$ 0.53 & 4.00 $\pm$ 2.29 & 3.55 $\pm$ 2.03 & 3.26 $\pm$ 1.45 & 3.04 $\pm$ 1.35 \\
$[$O~\textsc{ii}$]$ & 7329.66 & 1.71 $\pm$ 0.19 & 1.56 $\pm$ 0.17 & 2.19 $\pm$ 0.68 & 1.94 $\pm$ 0.60 & 1.99 $\pm$ 0.13 & 1.86 $\pm$ 0.12 \\
$[$Cl~\textsc{iv}$]$ & 7530.54 & - & - & - & - & - & - \\
$[$Ar~\textsc{iii}$]$ & 7751.06 & 1.36 $\pm$ 0.27 & 1.23 $\pm$ 0.24 & 1.34 $\pm$ 0.57 & 1.18 $\pm$ 0.50 & 1.74 $\pm$ 0.89 & 1.61 $\pm$ 0.82 \\
$[$Cl~\textsc{iv}$]$ & 8045.62 & - & - & - & - & - & - \\
$[$S~\textsc{iii}$]$ & 9069.00 & 7.81 $\pm$ 0.46 & 6.85 $\pm$ 0.40 & 8.26 $\pm$ 0.74 & 7.00 $\pm$ 0.63 & 9.88 $\pm$ 0.69 & 8.96 $\pm$ 0.63 \\
\midrule
F(H$\beta$) & - & 1.18 $\pm$ 0.03 & 1.46 $\pm$ 0.03 & 0.70 $\pm$ 0.02 & 0.92 $\pm$ 0.02 & 0.75 $\pm$ 0.02 & 0.88 $\pm$ 0.02 \\
$E(B-V)$ & - & 0.10 $\pm$ 0.02 & - & 0.12 $\pm$ 0.03 & - & 0.08 $\pm$ 0.03 & - \\
    \bottomrule
    \end{tabular}
\end{sidewaystable}

\begin{deluxetable*}{lcccccccc}[h]
%\digitalasset
\tablewidth{0pt}
\tablecaption{Ionic temperatures and abundances of the individual H~\textsc{ii} regions. \label{tab:individual_hii_region_abundances}}
\tablehead{
\colhead{Parameter} & \colhead{Region 1} & \colhead{Region 2} & \colhead{Region 3} & \colhead{Region 4} & \colhead{Region 5} & \colhead{Region 6} & \colhead{Region 7}
}
    \startdata
\hline
$n_e([$S~\textsc{ii}$])$ $[$cm$^{-3}]$ & 189.87 $\pm$ 78.70 & 60.92 $\pm$ 39.48 & 35.98 $\pm$ 27.59 & 61.36 $\pm$ 49.80 & 67.36 $\pm$ 56.55 & 85.04 $\pm$ 71.92 & 80.29 $\pm$ 58.13 \\
$T_e([$S~\textsc{iii}$])$ $[$10$^3$ K$]$ & 17.63 $\pm$ 0.81 & 20.46 $\pm$ 0.64 & 20.91 $\pm$ 2.41 & 16.55 $\pm$ 2.55 & 22.31 $\pm$ 4.80 & 20.02 $\pm$ 9.08 & 17.38 $\pm$ 2.76 \\
$T_e$(high) $[$10$^3$ K$]$ & 17.24 $\pm$ 0.76 & 19.91 $\pm$ 0.60 & 20.33 $\pm$ 2.27 & 16.22 $\pm$ 2.41 & 21.65 $\pm$ 4.53 & 19.49 $\pm$ 8.57 & 17.00 $\pm$ 2.61 \\
$T_e$(low) $[$10$^3$ K$]$ & 13.46 $\pm$ 0.35 & 14.69 $\pm$ 0.28 & 14.88 $\pm$ 1.04 & 13.00 $\pm$ 1.11 & 15.49 $\pm$ 2.08 & 14.50 $\pm$ 3.93 & 13.35 $\pm$ 1.20 \\
\midrule
log(O$^+$/H$^+$) & -5.17 $\pm$ 0.08 & -5.26 $\pm$ 0.05 & -4.97 $\pm$ 0.13 & -4.53 $\pm$ 0.20 & -5.07 $\pm$ 0.26 & -4.62 $\pm$ 0.54 & -4.68 $\pm$ 0.23 \\
log(O$^{2+}$/H$^+$) & -4.32 $\pm$ 0.05 & -4.51 $\pm$ 0.03 & -4.62 $\pm$ 0.10 & -4.77 $\pm$ 0.16 & -4.65 $\pm$ 0.17 & -4.41 $\pm$ 0.43 & -4.50 $\pm$ 0.16 \\
log(N$^+$/H$^+$) & -6.80 $\pm$ 0.04 & -6.84 $\pm$ 0.02 & -6.53 $\pm$ 0.06 & -6.33 $\pm$ 0.10 & -6.55 $\pm$ 0.12 & -6.43 $\pm$ 0.25 & -6.25 $\pm$ 0.10 \\
log(S$^+$/H$^+$) & -7.02 $\pm$ 0.04 & -6.99 $\pm$ 0.02 & -6.70 $\pm$ 0.06 & -6.42 $\pm$ 0.08 & -6.68 $\pm$ 0.11 & -6.51 $\pm$ 0.21 & -6.45 $\pm$ 0.08 \\
log(S$^{2+}$/H$^+$) & -6.36 $\pm$ 0.06 & -6.48 $\pm$ 0.02 & -6.53 $\pm$ 0.08 & -6.39 $\pm$ 0.11 & -6.51 $\pm$ 0.12 & -6.30 $\pm$ 0.28 & -6.28 $\pm$ 0.12 \\
log(Ar$^{2+}$/H$^+$) & -6.87 $\pm$ 0.06 & -6.97 $\pm$ 0.02 & -7.03 $\pm$ 0.08 & -6.98 $\pm$ 0.16 & -6.98 $\pm$ 0.14 & -6.81 $\pm$ 0.34 & -6.80 $\pm$ 0.19 \\
log(Ar$^{3+}$/H$^+$) & -6.97 $\pm$ 0.05 & -7.50 $\pm$ 0.06 & -7.39 $\pm$ 0.19 & - & -7.08 $\pm$ 0.43 & - & - \\
log(Cl$^{2+}$/H$^+$) & -8.21 $\pm$ 0.05 & -8.28 $\pm$ 0.06 & -8.75 $\pm$ 0.36 & - & -8.58 $\pm$ 0.44 & - & - \\
log(Cl$^{3+}$/H$^+$) & -8.10 $\pm$ 0.07 & - & - & - & - & - & - \\
\midrule
log(ICF(N$^+$)) & 0.90 $\pm$ 0.04 & 0.82 $\pm$ 0.03 & 0.51 $\pm$ 0.03 & 0.20 $\pm$ 0.03 & 0.55 $\pm$ 0.07 & 0.39 $\pm$ 0.16 & 0.41 $\pm$ 0.08 \\
log(ICF(S$^+$ + S$^{2+}$)) & 0.14 $\pm$ 0.01 & 0.12 $\pm$ 0.01 & 0.05 $\pm$ 0.01 & 0.006 $\pm$ 0.002 & 0.06 $\pm$ 0.01 & 0.03 $\pm$ 0.03 & 0.03 $\pm$ 0.01 \\
log(ICF(Ar$^{2+}$ + Ar$^{3+}$)) & 0.0094 $\pm$ 0.0002 & 0.0099 $\pm$ 0.0004 & 0.025 $\pm$ 0.003 & 0.12 $\pm$ 0.01 & 0.022 $\pm$ 0.005 & 0.08 $\pm$ 0.05 & 0.07 $\pm$ 0.01 \\
log(ICF(Cl$^{2+}$)) & -$^{a}$ & 0.21 $\pm$ 0.01 & 0.113 $\pm$ 0.005 & 0.13 $\pm$ 0.01 & 0.12 $\pm$ 0.01 & 0.11 $\pm$ 0.06 & 0.11 $\pm$ 0.01 \\
\midrule
12 + log(O/H) & 7.73 $\pm$ 0.05 & 7.56 $\pm$ 0.03 & 7.54 $\pm$ 0.11 & 7.67 $\pm$ 0.18 & 7.50 $\pm$ 0.19 & 7.84 $\pm$ 0.44 & 7.73 $\pm$ 0.17 \\
12 + log(N/H) & 6.10 $\pm$ 0.03 & 5.98 $\pm$ 0.03 & 5.99 $\pm$ 0.05 & 5.87 $\pm$ 0.10 & 6.01 $\pm$ 0.10 & 5.99 $\pm$ 0.27 & 6.16 $\pm$ 0.11 \\
12 + log(S/H) & 5.86 $\pm$ 0.05 & 5.75 $\pm$ 0.02 & 5.74 $\pm$ 0.07 & 5.90 $\pm$ 0.09 & 5.77 $\pm$ 0.10 & 5.95 $\pm$ 0.25 & 5.97 $\pm$ 0.10 \\
12 + log(Ar/H) & 5.39 $\pm$ 0.05 & 5.15 $\pm$ 0.03 & 5.16 $\pm$ 0.10 & 5.14 $\pm$ 0.16 & 5.27 $\pm$ 0.28 & 5.27 $\pm$ 0.34 & 5.27 $\pm$ 0.19 \\
12 + log(Cl/H) & 4.15 $\pm$ 0.06 & 3.93 $\pm$ 0.06 & 3.36 $\pm$ 0.36 & - & 3.54 $\pm$ 0.44 & - & - \\
\midrule
log(N/O) & -1.63 $\pm$ 0.05 & -1.58 $\pm$ 0.04 & -1.56 $\pm$ 0.08 & -1.79 $\pm$ 0.14 & -1.47 $\pm$ 0.16 & -1.81 $\pm$ 0.38 & -1.57 $\pm$ 0.17 \\
log(S/O) & -1.87 $\pm$ 0.02 & -1.80 $\pm$ 0.02 & -1.80 $\pm$ 0.05 & -1.76 $\pm$ 0.10 & -1.73 $\pm$ 0.09 & -1.89 $\pm$ 0.21 & -1.75 $\pm$ 0.10 \\
log(Ar/O) & -2.34 $\pm$ 0.02 & -2.41 $\pm$ 0.02 & -2.38 $\pm$ 0.05 & -2.53 $\pm$ 0.12 & -2.24 $\pm$ 0.24 & -2.58 $\pm$ 0.18 & -2.46 $\pm$ 0.16 \\
log(Cl/O) & -3.58 $\pm$ 0.03 & -3.63 $\pm$ 0.05 & -4.18 $\pm$ 0.35 & - & -3.96 $\pm$ 0.42 & - & - \\
    \enddata
    \tablecomments{$^{a}$Cl/H calculated using Eq.~\ref{eq:Cl}.}
\end{deluxetable*}

%% For this sample we use BibTeX plus aasjournalv7.bst to generate the

\bibliography{um461}{}
\bibliographystyle{aasjournalv7}



\end{document}